\documentclass[review]{elsarticle}

\usepackage[ruled,vlined]{algorithm2e}
\usepackage[breaklinks]{hyperref}
\usepackage{threeparttable}
\usepackage{mathtools}
\usepackage{microtype}
\usepackage{longtable}
\usepackage{pdflscape}
\usepackage{setspace}
\usepackage{tabularx}
\usepackage{multirow}
\usepackage{ragged2e}
\usepackage{textcomp}
\usepackage{booktabs}
\usepackage{amsmath}
\usepackage{amssymb}
\usepackage{lscape}
\usepackage{lineno}
\usepackage{csquotes}
\usepackage[table]{xcolor}
\usepackage{array}
\usepackage{float}
\usepackage{tabu}
\usepackage{todonotes}
\usepackage{hyperref}
\hypersetup{
colorlinks=true,
linkcolor=blue,
filecolor=blue,
urlcolor=blue,
citecolor=blue,
}

\usepackage{geometry}
\modulolinenumbers[1]

\journal{Engineering Applications of Artificial Intelligence}

\biboptions{numbers,sort&compress}
\begin{document}
\begin{frontmatter}
\title{Automated liver tissues delineation techniques: A systematic survey on machine learning current trends and future orientations}

\author[QU_EE]{Ayman Al-Kababji}\corref{correspondingauthor}
\cortext[correspondingauthor]{Corresponding author}
\ead{aa1405810@qu.edu.qa}

\author[QU_EE]{Faycal Bensaali}
\ead{f.bensaali@qu.edu.qa}

\author[HMC]{Sarada Prasad Dakua}
\ead{SDakua@hamad.qa}

\author[QU_EE]{Yassine Himeur}
\ead{yassine.himeur@qu.edu.qa}

\address[QU_EE]{Department of Electrical Engineering, Qatar University, Doha, Qatar}
\address[HMC]{Department of Surgery, Hamad Medical Corporation, Doha, Qatar}

\begin{abstract}
Machine learning and computer vision techniques have grown rapidly in recent years due to their automation, suitability, and ability to generate astounding results. Hence, in this paper, we survey the key studies that are published between 2014 and 2022, showcasing the different machine learning algorithms researchers have used to segment the liver, hepatic tumors, and hepatic-vasculature structures. We divide the surveyed studies based on the tissue of interest (hepatic-parenchyma, hepatic-tumors, or hepatic-vessels), highlighting the studies that tackle more than one task simultaneously. Additionally, the machine learning algorithms are classified as either supervised or unsupervised, and they are further partitioned if the amount of work that falls under a certain scheme is significant. Moreover, different datasets and challenges found in literature and websites containing masks of the aforementioned tissues are thoroughly discussed, highlighting the organizers' original contributions and those of other researchers. Also, the metrics used excessively in literature are mentioned in our review, stressing their relevance to the task at hand. Finally, critical challenges and future directions are emphasized for innovative researchers to tackle, exposing gaps that need addressing, such as the scarcity of many studies on the vessels’ segmentation challenge and why their absence needs to be dealt with sooner than later.
\end{abstract}

\begin{keyword}
Liver \sep hepatic-tumors \sep hepatic-vessels \sep machine learning \sep survey \sep semantic segmentation
\MSC[2010] 00-01\sep 99-00
\end{keyword}

\end{frontmatter}

\section{Introduction}\label{sec1}

Two million deaths annually around the world are credited to hepatic diseases \cite{Asrani2019}. Half of these deaths are related to complications caused by liver cirrhosis, and the other half are due to hepatitis and hepatocellular carcinoma (HCC) \cite{Asrani2019}. Unfortunately, it is also a hub for metastasis originating from adjacent organs such as the colon, rectum, pancreas, stomach, esophagus, breasts, lungs, etc. \cite{Surgery2020}. Regardless of the tumors' origin, the liver and its lesions are routinely analyzed in primary tumor staging \cite{Christ2017}. In particular, HCC comprises a genetically and molecularly heterogeneous group of cancers commonly arising in chronically damaged livers \cite{Christ2017}. However, screening for liver-related diseases can reduce mortality \cite{NASIRI_2020}. 
Early detection and accurate delineation of hepatic tumors can help a physician decide on more appropriate treatment planning.

\begin{center}
\begin{tabular}{|m{17cm}|}
\hline
{\small \textbf{Abbreviations:}} \\ 
{\scriptsize 
\color{black}
2-dimensional: 2D;
3-dimensional: 3D;
3D anisotropic hybrid network: 3D AH-Net;
3D conditional random field: 3D CRF;
3D deformable model optimization: 3D DMO;
3D Image Reconstruction for Comparison of Algorithm Database: 3D-IRCADb;
3D universal U-net: 3D U\textsuperscript{2}-Net;
active shape models: ASM;
adaptive boosting: AdaBoost;
adaptive radiation therapy: ART;
artificial neural network: ANN;
area under curve: AUC;
artificial intelligence: AI;
attention hybrid connection blocks network: AHCNet;
attention mechanism and nested U-Net: ANU-Net;
average symmetric surface distance: ASD;
cascaded 2D FCN: CFCN;
cascaded conditional GANs: CCGAN;
cascaded random forest: CaRF;
cascaded U-ResNet: CU-ResNet;
Combined (CT-MR) Healthy Abdominal Organ Segmentation: CHAOS;
complementary network: CompNet;
computer-aided detection: CAD;
computerized tomography: CT;
contrast-enhanced CT: CE-CT;
convolutional neural network: CNN;
convolutional long short-term memory: C-LSTM;
convolutional denoising autoencoders: CdAE;
conjugate FCN: CoFCN;
CT Volumes with Multiple Organ Segmentations: CT-ORG;
decision tree: DT;
deep atlas prior: DAP;
deep belief network: DBN;
deep data-driven loss: DDL;
deep image-to-image network: DI2IN;
deep learning: (DL);
deeply supervised network: DSN';
densely-connected generative adversarial network: DC-GAN;
densely-connected U-Net: DenseU-Net;
digitally reconstructed radiographs: DRR;
Dice Similarity Coefficient: DSC;
Dice/Case: DPC;
domain adaptation: DA;
dual-branch progressive: DBP;
dynamic contrast-enhanced: DCE;
extreme learning machine: ELM;
false positive: FP;
false negative: FN;
false positive rate: FPR;
fast fuzzy c-means: FFCM;
fast kernelized fuzzy c-means: FKFCM;
finite element method: FEM;
fractal residual network: FRN;
fully convolutional network: FCN;
fuzzy c-means: FCM;
fuzzy connectedness: FC;
gastrointestinal: GI;
generative adversarial network: GAN;
genetic optimizer: GO;
global and local context U-Net: GLC-UNet;
global DSC: GDSC;
graph attention layers: GAT;
graph-cut: GC;
graph convolutions-based ResNet: G-ResNet;
graph neural network: GNN;
grey wolf optimization: GWO;
Hausdorff Distance: HD;
hepatocellular carcinoma: HCC;
holistic nested network: HNN;
hybrid DenseU-Net: H-;
IEEE International Symposium on Biomedical Imaging: ISBI;
Imaging Methods Assessment and Reporting: IMAR;
intersection-over-union: IoU;
kernelized FCM: KFCM;
laplacian forest: LF;
laws texture energy measure: LTEM;
level-set methods: LSM;
liver extraction residual convolutional network: LER-CN;
Liver Tumor Segmentation Challenge: LiTS;
local direction of gradient: LDOG;
Jaccard index: JI;
machine learning: ML;
magnetic resonance imaging: MRI;
Markov random field: MRF;
maximum symmetric surface distance: MSD;
mean shape fitting: MSF;
Medical Image Computing and Computer Assisted Intervention: MICCAI;
Medical Segmentation Decathlon Challenge: MSDC;
MSDC Task 3: MSDC-T3;
MSDC Task 8: MSDC-T8;
MIDAS Liver Tumor: MIDAS-LT;
modified U-Net: mU-Net;
Multi-Atlas Labeling Beyond the Cranial Vault: BtCV;
multi-channel FCN: MC-FCN;
multi-channel 3D FCN ResNet: MC-FC-ResNet;
multi-planar network: MPNet;
multi-planar U-Net: MPU-Net;
multiple-input and multiple-output feature abstraction network: MIMO-FAN;
multi-scale candidate generation: MCG;
National Library of Medicine: NLM;
neutrosophic sets: NS;
noise removal component: NRC;
non-negative matrix factorization: NMF;
particle swarm optimization: PSO;
positive predictive value: PPV;
positron-emitting computerized tomography: PET-CT;
prior-aware neural network: PaNN;
project and excite: PE;
random forests: RF;
random walks: RW;
receiver operating characteristic: ROC;
relative volumetric distance: RVD;
relaxed upper confident bound: RUCB;
residual attention-aware U-Net: RA-UNet;
response evaluation criteria in solid tumors: RECIST;
root-mean-square symmetric surface distance: RMSD;
robust features: SURF;
scalable key points: BRISK;
Segmentation of the LIVER Competition 2007: SLIVER07;
selective internal radiation therapy: SIRT;
signal-to-noise ratio: SNR;
simple linear iterative clustering: SLIC;
squeeze and channel excitation: cSE;
statistical shape models: SSM;
structural preservation component: SPC;
support vector machine: SVM;
task-driven generative adversarial network: TD-GAN;
The Cancer Imaging Archive: TCIA;
tissue of interest: TOI;
transfer learning: TL; 
true positive: TP;
true negative: TN;
true Positive Rate: TPR;
true Negative Rate: TNR;
two-path CNN: TPCNN;
U-Net ResNet: U-ResNet;
ultrasound: US;
variational autoencoders: VAE;
Vascular Synthesizer: VascuSynth;
Visual Concept Extraction Challenge in Radiology: VISCERAL;
volume attention Mask-RCNN: VA Mask-RCNN;
volumetric overlap error: VOE;
} \\ \hline
\end{tabular}
\end{center}

\textcolor{black}{As important as the liver is, many developed imaging modalities such as computerized tomography (CT), magnetic resonance imaging (MRI), positron-emitting computerized tomography (PET-CT), and ultrasound (US) are used for the liver’s morphological and volumetric analysis and diagnosis of associated diseases \cite{Campadelli2009,Gotra2017}. These modalities are deemed to be useful, especially for their capability of giving surgeons insights into the current state of organs non-invasively. With the existence of such modalities, computer-aided detection (CAD) systems have become significantly important. Furthermore, CT, MRI, and PET can generate 2-dimensional (2D) slices of the human body, which can be combined to generate 3D holistic organ volumes for surgeons to analyze. Thus, they bear more advantages than the US modality, especially in providing clearer and more informative image slices. Moreover, thanks to CT scans’ higher signal-to-noise ratio (SNR) and better spatial resolution, they produce more accurate anatomical information about the visualized structures, and this imaging technique is preferred by diagnosticians \cite{Campadelli2009}. Moreover, relative to MRI, CT scans have a shorter acquisition time \cite{Gotra2017}. In contrast, the patient is more exposed to radiation in modalities like CT. Additionally, the chances of developing fatal cancers from CT scans are 1 in 2,000, which is fairly small; however, with the increased number of scans, the chances become higher \cite{Chang2018}.}

CT, MRI, and PET-CT allow clinicians, physicians, and surgeons to have a clearer insight into the body organs non-invasively. CT scans, for instance, provide three different anatomical views for the organs from transversal, sagittal and coronal planes, allowing medical personnel to tackle the organ of interest from different views. Such modalities are utilized extensively by medical personnel for countless clinical applications, including organic cancer diagnosis, organ transplantation, and surgical planning \cite{Wang2018}. All these procedures apply in the case of the liver, where different cancerous cells exist, such as HCC, cysts, metastases, etc. Additionally, such modalities are used for adaptive radiation therapy (ART), which is a radiation treatment plan that imposes modifications based on the patient's functional changes during a course of radiation \cite{Liang2018}. In another clinical procedure, a pre-procedural CT or MRI scan can help in interventional endoscopy for pancreatic and biliary diseases as image guidance can be supportive in intra-procedural navigation \cite{Gibson2018} to specific gastrointestinal (GI) positions as the endoscope’s field of view is small and lacks visual orientation cues \cite{Gibson2017}. \textcolor{black}{Also, medical image registration can aid medical practitioners in observing the motion of intra-patient organs in the middle of procedures \cite{Heinrich2019}. Thus, it is evident how important these modalities are in increasing the quality of life and expectancy for numerous patients.}

\textcolor{black}{All the aforementioned reasons justify the idea of segmenting humans' organs, especially the liver, its tumors, and vessels, to aid medical personnel in disease diagnosis. Not only is it important to segment liver and tumors pre-procedural, but numerous advantages also prevail when performing post-procedural segmentation on a follow-up CT scan. For instance, through medical image registration, a realization can be achieved about whether the conducted procedure was successful in stopping the disease or not.}
Furthermore, tumor burden quantification, which measures the volume of all tumors within the liver \cite{Vivanti2017}, is important when discussing tumors' progression within the liver. Early diagnosis and accurate segmentation of hepatic tumors can help doctors plan an appropriate treatment procedure \cite{NASIRI_2020}. Additionally, a follow-up CT scan, which segments the liver and the tumors, is also of interest since diseases' progression can be documented for further analysis and treatment procedure planning. However, the norm currently in clinical routines is to manually or semi-automatically segment the liver from CT and MRI modalities. Even though, in some scenarios, these techniques can be more accurate than automatic ones \cite{Zheng2017}, the underlying issues of manual and semi-automatic techniques are represented by their subjectivity (i.e., dependency on the radiologists' experience), intra- and inter-radiologist variance, and time-consumption \cite{Hu2016}, especially for experts whose time is extremely valuable. Thus, comes the importance of using automatic methods with high segmentation performance.

Many devised automatic segmentation techniques have been applied in the last two decades. They can be categorized into statistical-based and learning-based approaches, where the former can be represented by scans intensities' statistical distribution, including atlases, statistical shape models (SSM), active shape models (ASM), level-set methods (LSM), and graph-cut (GC) methods \cite{Yang2017}. Usually, these methods are challenged by boundary leakage, under- or over-segmentation \cite{Zheng2017}. The latter, on the other hand, depends on either hand-crafted features as in conventional machine learning (ML) algorithms or empirically-found features as in the case of convolutional neural network (CNN), \textcolor{black}{which is a special structure of the artificial neural network (ANN)}. The medical image segmentation field made the most significant leap riding on the wave of deep CNNs \cite{Wang2018}, where they reached a state of capability that enabled them to generate expert-like segmentations in extremely minimal time. Hence, ML techniques are essential and effective tools in analyzing highly complex data from the medical fields \cite{garg2021role}.

However, creating an accurate segmentation of the liver, hepatic tumors, and hepatic vessels is still a challenge. During data acquisition, the variance within the dataset constructs obstacles in front of the model's performance, such as the different scanning protocols with different voxel densities/scanners resolution and different contrast agents with varying levels of contrast enhancements \cite{Christ2017}. On the other hand, from an organs point of view, the low-contrast boundaries exhibited between the liver and surrounding organs create areas of fuzziness that are hard for models to classify, which are translated into over- or under-segmentation \cite{Hu2016}. Moreover, the highly varying liver shapes and/or sizes among people, especially abnormalities introduced by surgical resection \cite{Christ2017}, make it harder for liver segmentation techniques, particularly for SSM and similar predictors \cite{Hu2016}. Challenges of segmenting what is within the liver are introduced by the heterogeneity of tumors' sizes and shapes, and intra-hepatic veins (vessels) irregularities, which further complicates the segmentation task \cite{Hu2016,Christ2017}. Thus, creating a detailed 3D liver segmentation detailing the exact components is one of the most challenging tasks that need addressing.

Similar to what \cite{garg2021role} did, but with a more in-depth review of the ML techniques focusing on the liver, this paper highlights their use in the liver's segmentation from medical images (including the tumors and vessels) by:
\begin{itemize}
\item Reviewing the state-of-the-art algorithms incorporating/based on novel ML
\item Specifying all the publicly-available datasets' for liver, tumors, and/or vessels segmentation challenge
\item Highlighting the metrics currently in use for evaluating the model's segmentation performance
\item Presenting the literature based on ML to segment liver, tumors, and/or vessels during 2014 - 2022
\item Providing a list of open research issues and future directions for improving the available existing datasets and the automatic segmentation techniques
\end{itemize}

The remainder of this paper is organized as follows. Section \ref{sec2} highlights the existing datasets, which can be utilized for the aforementioned cause. Section \ref{sec3} describes the objective metrics that evaluate created models for performance comparison. Section \ref{sec4} delves into the related works, where a comparison of compartments is shown and discussed. Section \ref{sec5} talks about the current challenges and future directions we observed when investigating the related works, and finally, we conclude the paper in Section \ref{sec6}.

\section{Publicly Available Challenges and Datasets}\label{sec2}
This section provides a brief historical background and a summary of the literature's datasets/challenges' specifications. Each dataset's origins and specifications are discussed and then summarized in Table~\ref{tab:datasets_comparison}. Some of the mentioned datasets/challenges were hosted by well-known medical conferences such as the Medical Image Computing and Computer Assisted Intervention (MICCAI) conference and the IEEE International Symposium on Biomedical Imaging (ISBI). The organizers would present the datasets as challenges pushing researchers to participate by creating healthy peer-pressure environments. Other datasets were shared publicly by different research institutions to push researchers forward to create better algorithms\footnote{Some of the datasets were investigated with the aid of ITK-SNAP \cite{ITK-SNAP}. Available at: www.itksnap.org}.

\subsection{Inclusion/Exclusion Criteria}
This paper comprehensively highlights and presents datasets that include liver, tumors, and/or vessel delineations, solely or with other organs' ground-truth masks.

However, if the ground-truth labels for either liver, tumors, or vessels were not created, we refrain from including that particular dataset. Lastly, we define ground-truth labels as tissues' delineation (i.e., segmentation), not the localization of said tissues.

\subsection{Datasets/Challenges}
\subsubsection{Segmentation of the LIVER Competition [2007] (SLIVER07)}
The competition occurred in a workshop named ``3D Segmentation in the Clinic: A Grand Challenge'' in October in conjunction with MICCAI 2007. The results of that workshop are summarized in \cite{Heimann2009}. The dataset has 30 contrast-enhanced CT (CE-CT) scans divided into 20 training volumes and 10 testing volumes. The intra-slice resolution varies between 0.54 and 0.86 mm, while the inter-slice space varies between 0.5 and 5 mm. The number of pixels is the same for all the volumes (512$\times$512), with varying slices ranging between 64 and 502.

\subsubsection{3D Image Reconstruction for Comparison of Algorithm Database (3D-IRCADb) [\texorpdfstring{$\leq$}{≤} 2010]}
3D-IRCADb is a database gathered by the IRCAD institute in France, which includes anonymized medical images of patients. In total, the dataset has 22 venous phase CE-CT scans divided into: 1) 3D-IRCADb01, which contains 10 males and 10 females with 75\% having hepatic tumors; 2) 3D-IRCADb-02, which contains 2 CT scans with other abdominal organs segmented. The intra-slice resolution for the whole dataset varies between 0.56 and 0.96 mm, while the inter-slice distance varies between 1 and 4 mm. On the other hand, (512 $\times$ 512) pixels are used per slice, while the number of slices ranges between 74 and 260. It is worth noting that the majority of literature focuses on the 3D-IRCADb01 part and is normally divided into training and testing records accordingly.

\subsubsection{MIDAS Liver Tumor (MIDAS-LT) Segmentation Dataset [2010]}
MIDAS-LT, an acronym we created, is a part of a bigger initiative to provide a collection of archived, analyzed, and publicly accessed datasets called MIDAS \cite{MIDAS}. The MIDAS-LT is funded by the National Library of Medicine (NLM) in the USA under the Imaging Methods Assessment and Reporting (IMAR) project. The dataset contains 4 CT scans with (up to) 3 radiologists' manual segmentation for liver tumors per volume, without a mask of the liver. All the dimensions (inter- and intra-slice) vary between 1.73 and 1.85 mm. On the other hand, the number of pixels per slice varies between 177 and 189, while the number of slices is between 98 and 259. It is worth noting that the original dataset had more homogeneous specifications, but the dimensions reported here are for the segmented volumes.

\subsubsection{Vascular Synthesizer (VascuSynth) [2013]} 
VascuSynth is a software for synthesizing tubular-shaped structures such as human organ vessels or other tree-like structures. Due to the absence of datasets with manually segmented vessels for training, the creators \cite{hamarneh2010vascusynth,jassi2011vascusynth} aimed to create synthesized data to support the cause of automated segmentation of tubular structures in 3D medical images. The software is capable of simulating volumetric vascular images by iteratively growing vascular trees based on either user-defined or spatially varying oxygen demand maps. Moreover, the software generates the corresponding ground truth segmentations, the tree hierarchy, the bifurcation locations, and the branch properties. In 2013, they created 120 vascular volumes, divided into 10 groups of 12 records each as a showcase for the software capabilities. We have not added it to the summary table because it is synthesized and thus, can be generated per the user/researcher's requirements.

\subsubsection{Multi-Atlas Labeling Beyond the Cranial Vault (BtCV) - Workshop and Challenge [2015]}
This workshop is the last of a series of workshops introduced by Landman et al. from Vanderbilt University \cite{BtCV} hosted by MICCAI. Their aim in this workshop is to extend multi-atlas segmentation beyond the skull vault to include the cervix and abdomen segmentation. Thus, the liver, among other organs (kidneys, gallbladder, esophagus, stomach, etc.) is segmented. The dataset contains many segmentations and registrations, but the number of liver segmentation records is 50 venous phase CE-CT scans, divided into 30 training and 20 testing records found under the `RawData' file. The intra-slice resolution is between 0.54 and 0.98 mm, while the inter-slice distance is between 2.5 to 5 mm. In contrast, the number of pixels is 512$\times$512, with the number of slices varying between 85 and 198.

\subsubsection{Pancreas-CT [2015]}
Pancreas-CT is a portal venous CE-CT dataset that contains pancreas manual delineations available on The Cancer Imaging Archive (TCIA) website \cite{Pancreas-CT}. It originally contained 82 records when first published in 2015. However, in 2020, 2 records were removed (\#25 \& \#70) as they were duplicates of Record \#2 with slight variations. For the voxels' physical dimensions, the intra-slice resolution varies between 0.66 and 0.98 mm, while the inter-slice distance ranges between 0.5 to 1 mm. On the other hand, the number of pixels is 512$\times$512, with the number of slices being between 181 and 466. The reason behind including this dataset is that the liver delineations are created along with other organs for 43 records (42 after removing the duplicate Record~\#25) from this dataset \cite{Gibson2018}. It is worth noting that the specifications mentioned in Table~\ref{tab:datasets_comparison} are for those records with liver delineation provided in \cite{Gibson2018}. These records are representative of all the dimensions of the original dataset; however, the minimum number of slices is 186 instead.

\subsubsection{Visual Concept Extraction Challenge in Radiology (VISCERAL) Anatomy3 [2016]}
The challenge took place for three consecutive years (2014 - 2016) in conjunction with ISBI. The challenge is concerned with multi-organ segmentation providing ground-truth labeling of up to 20 organs (liver, pancreas, spleen, kidneys, lungs, aorta, urinary bladder, gallbladder, etc.). The full list can be found in \cite{VISCERAL}, and the results of all the workshops are summarized in \cite{Toro2016}. The dataset has 120 records, from CT and MRI modalities, with and without contrast-enhancing agents (refer to Table~\ref{tab:datasets_comparison} for further details). For the CT records, the intra-slice resolution varies between 0.60 and 1.40 mm, while the inter-slice distance is fixed to 3 mm \cite{Toro2016}. For the MRI, the slice resolution is between 0.84 and 1.30 mm, while the inter-slice distance varies between 3 and 8 mm \cite{Toro2016}.

\subsubsection{Liver Tumor Segmentation Challenge (LiTS) [2017]}
This challenge was conducted in both ISBI (18/04/2017) and MICCAI (14/09/2017) to provide researchers with ground-truth labels for the liver and tumors within. The challenge is to automatically segment liver tumors/lesions in CT volumes and estimate tumors' burden, along with the typical liver segmentation challenge. The dataset has 201 CE-CT records in total, divided into 131 training and 70 testing scans. The dataset can be found in \cite{LiTS}, and the summary of the challenge results is summarized in \cite{Patrick2019}. Noting that 3D-IRCADb01 is part of the training set of LiTS (Records 28 - 47 \cite{Jiang2019}), care must be taken when both 3D-IRCADb and LiTS datasets are used to train the model. By doing this, the model will be biased towards the common records since the model is trained on them twice in every epoch. Also, it is inappropriate to train on the full training set of LiTS and test on 3D-IRCADb, as the testing set would be exposed to the model in the training phase beforehand. For the physical dimensions, the intra-slice resolution varies between 0.55 and 1.00 mm, while the inter-slice distance ranges between 0.45 to 6.0 mm \cite{Patrick2019}. In contrast, the number of pixels is 512$\times$512, with the number of slices ranging between 42 and 1026.

\subsubsection{Medical Segmentation Decathlon Challenge (MSDC) [2018]}
MSDC was held in MICCAI 2018, where it uniquely focuses on the segmentation generalizability of a model on 10 different biomedical tasks. In this review, we only report the liver-related tasks, which are Task 3 and Task 8; however, details regarding all the tasks are summarized in \cite{simpson2019large}. We refer to Task 3 and Task 8 datasets as MSDC-T3 and MSDC-T8, respectively.

As a matter of fact, MSDC-T3 is the same as the LiTS dataset, where the training sets are identical, but the testing set in MSDC-T3 is shuffled when compared to its counterpart in LiTS. On the other hand, the MSDC-T8 dataset contains 443 portal venous phase CE-CT scans with segmented tumors and vessels only, where 303 are designated as training and the remaining 140 records as testing. The intra-slice resolution varies between 0.56 and 0.98 mm, while the inter-slice distance ranges between 0.80 and 8 mm. The slices have the standard number of pixels for a CT scan, which is (512 $\times$ 512), with varying slices between 24 and 251. It is worth noting that the authors in \cite{Tian2019} created the liver annotations within the MSDC-T8 443 CT records and shared them publicly. Couinaud's segmentation of 193 livers among the 443 records is also shared.

\subsubsection{CT Volumes with Multiple Organ Segmentations (CT-ORG) Dataset [2019]}
CT-ORG is an extension of the LiTS dataset and is publicly accessible via the TCIA website \cite{TCIA}. It contains 140 CT scans where the creators \cite{CT-ORG} added extra 9 PET-CT scans over the LiTS training set and extended the segmentation to multiple organs (lungs, bones, liver, kidneys, bladder, and brain). The majority of provided segmentations are golden-corpus (manually labeled), while lungs and bones in the training set are silver-corpus (automatically segmented). It is not mentioned whether the new records/organs have their tumors segmented or not. For the voxels' physical dimensions, the intra-slice resolution varies between 0.55 and 1.37 mm, while the inter-slice distance ranges between 0.7 to 5 mm \cite{Patrick2019}. On the other hand, the number of pixels is 512$\times$512 with varying slices between 74 and 987. The difference between LiTS dimensions mentioned earlier, and CT-ORG is contributed to the testing set of LiTS, which is not included in CT-ORG. We have verified this difference by developing a Python code to find the minimum and maximum of each quantity in both datasets.

\subsubsection{Combined (CT-MR) Healthy Abdominal Organ Segmentation (CHAOS) Challenge [2019]}
CHAOS was held in ISBI 2019, aiming to segment abdominal multi-organ tumor-free CT and MRI data. The dataset has both CT and MRI (T1 and T2 weighted) parts, where there is no inter-modality connection (i.e., the CT and MRI data are from random patients, not counterparts for the same patient). The summary of this challenge is reported in \cite{kavur2020chaos}.

The CT dataset contains 40 CE-CT records for patients with the ground-truth label for healthy livers (potential liver donors) acquired at the portal venous phase. The intra-slice resolution varies between 0.7 and 0.8 mm, while the inter-slice space varies between 3 and 3.2 mm \cite{CHAOS}. The resolution is similar to other datasets (512$\times$512), and the number of slices ranges between 77 and 105 \cite{CHAOS}. On the other hand, the MRI dataset has ground-truth labeling for the liver, kidneys, and spleen, containing 120 records of both 80 T1-Dual (40 in-phase and 40 out-phase) and 40 T2-SPIR weighted records. The MRI records from different enhancing protocols are for the same patient. For instance, patient 20 has three MRI records falling into the three previously mentioned categories. Intra-slice resolution varies between 1.36 and 1.89 mm, while the inter-slice distance is between 5.5 and 9 mm \cite{CHAOS}. It is worth noting that the resolution here is different (256$\times$256), and the number of slices varies between 26 and 50 \cite{CHAOS}. The intra-modality acquisition protocol in this dataset is consistent where we see minor variations between different records belonging to the same modality. Further details can be found in \cite{CHAOS} that include the used acquisition devices and the different contrast enhancing phases for MRI.

\subsection{Comparison and Summary of Challenges/Datasets}
In Table~\ref{tab:datasets_comparison}, each dataset's positives and negatives are highlighted explicitly. It can be noticed that the most comprehensive dataset (from a liver point of view) is the MSDC-T8 due to the presence of many records in that dataset, along with the vessels' segmentation availability.
Moreover, a summary of the reviewed challenges and datasets is depicted in this table. It highlights the website from where researchers can retrieve the datasets, the type of modality used, and the inclusion of contrast enhancement agents. Also, it highlights the training/testing ratio along with the physical and computerized dimensions for each dataset.

\begin{landscape}
\begin{table}[H]
\centering
\scriptsize
\tabulinesep=1.2mm
\caption{\textcolor{black}{Summary of available datasets and their characteristics.}} \label{tab:datasets_comparison} \vspace{2em}
\begin{tabu}{
|>{\centering\arraybackslash}m{1.5cm}
|>{\centering\arraybackslash}m{0.9cm}
|>{\centering\arraybackslash}m{1.2cm}
|>{\centering\arraybackslash}m{1.1cm}
|>{\centering\arraybackslash}m{1.5cm}
|>{\centering\arraybackslash}m{4.4cm}
|>{\centering\arraybackslash}m{3.3cm}
|m{3.8cm}
|m{2.8cm}
|}
\hline

Dataset & \textcolor{black}{Year} & Available Masks & Modality & \textcolor{black}{Size (Train/Test)} & Voxels Dimensions (Height~$\times$~Width~$\times$~Depth mm\textsuperscript{3}) & Volumes Dimensions (Height~$\times$~Width~$\times$~Slice) & Positives & Negatives \\ \hline

SLIVER07 \cite{SLIVER07} & \textcolor{black}{2007} & Liver & CE-CT & \textcolor{black}{30 (20/10)} & (0.54 $\sim$ 0.86) $\times$ (0.54 $\sim$ 0.86) $\times$ (0.5 $\sim$ 5) & 512 $\times$ 512 $\times$ (64 $\sim$ 502) & - The earliest publicly available dataset to have liver masks & - Small size ($\downarrow$) \newline - Does not have liver tumors and vessels \\ \hline

3D-IRCADb \cite{IRCAD} & \textcolor{black}{\texorpdfstring{$\leq$}{≤} 2010} & Liver (Tumors, Vessels) & CE-CT & \textcolor{black}{22~(N/A)} & (0.56 $\sim$ 0.96) $\times$ (0.56 $\sim$ 0.96) $\times$ (1 $\sim$ 4) & 512 $\times$ 512 $\times$ (74 $\sim$ 260) & - First dataset to include liver tumors \newline - Records metadata are mentioned & - Small size ($\downarrow\downarrow$) \newline - The Majority does not have liver vessels \\ \hline

MIDAS-LT \cite{MIDAS-LT} & \textcolor{black}{2010} & Tumors in Liver & CT & \textcolor{black}{4~(N/A)} & (1.73 $\sim$ 1.85) $\times$ (1.73 $\sim$ 1.85) $\times$ (1.73 $\sim$ 1.85) & (177 $\sim$ 189) $\times$ (177 $\sim$ 189) $\times$ (98 $\sim$ 259) & - N/A & - Very small size ($\downarrow\downarrow\downarrow$) \newline - Very few segmented tumors \\ \hline

BtCV \cite{BtCV} & \textcolor{black}{2015} & Liver \& Others & CE-CT & \textcolor{black}{50~(30/20)} & (0.54 $\sim$ 0.98) $\times$ (0.54 $\sim$ 0.98) $\times$ (2.5 $\sim$ 5) & 512~$\times$~512~$\times$~(85~$\sim$~198) & - Many organs segmented & - Medium size \newline - Does not have liver tumors and vessels \\ \hline

Pancreas-CT \cite{Pancreas-CT} & \textcolor{black}{2015} & Pancreas, Liver \& Others & CE-CT & \textcolor{black}{42~(N/A)} & (0.66 $\sim$ 0.98) $\times$ (0.66 $\sim$ 0.98) $\times$ (0.5 $\sim$ 1) & 512~$\times$~512~$\times$~(186~$\sim$~466) & - Has both pancreas and liver masks \newline - High resolution & - Medium size ($\uparrow$) \newline - Does not have liver tumors and vessels \\ \hline

\multirow{5.5}{1.5cm}{\centering VISCERAL Anatomy3 \cite{VISCERAL}} & \multirow{5.5}{*}{\centering\textcolor{black}{2016}} & \multirow{5.5}{1.2cm}{\centering Liver \& Others} & CT & \textcolor{black}{30 (20/10)} & (0.97 $\sim$ 1.40) $\times$ (0.97 $\sim$ 1.40) $\times$ 3 & \textbf{---------} & \multirow{5.5}{3.8cm}{- Multiple modalities \newline - Large size ($\uparrow\uparrow$) \newline - Highest number of segmented organs ($\sim$20 ground-truth organs)} & \multirow{5.5}{2.8cm}{- Hard to access \newline - Does not have liver tumors and vessels} \\ \cline{4-7}
    & & & CE-CT & \textcolor{black}{30 (20/10)} & (0.60 $\sim$ 0.79) $\times$ (0.60 $\sim$ 0.79) $\times$ 3 & \textbf{---------} & & \\ \cline{4-7}
    & & & MRI & \textcolor{black}{30 (20/10)} & 1.25 $\times$ 1.25 $\times$ 5 & \textbf{---------} & & \\ \cline{4-7} 
    & & & CE-MRI & \textcolor{black}{30 (20/10)} & (0.84 $\sim$ 1.30) $\times$ (0.84 $\sim$ 1.30) $\times$ (3 $\sim$ 8) & \textbf{---------} & & \\ \hline

LiTS \cite{LiTS} & \textcolor{black}{2017} & Liver (Tumors) & CE-CT & \textcolor{black}{201 (131/70)} & (0.55 $\sim$ 1.00) $\times$ (0.55 $\sim$ 1.00) $\times$ (0.45 $\sim$ 6) & 512~$\times$~512~$\times$~(42~$\sim$~1026) & - Large size ($\uparrow\uparrow$) \newline - Many tumors are segmented in 131 records & - Does not have liver vessels \\ \hline

MSDC-T8 \cite{MSDC} & \textcolor{black}{2018} & Liver (Tumors, Vessels) & CE-CT & \textcolor{black}{443 (303/140)} & (0.56 $\sim$ 0.97) $\times$ (0.56 $\sim$ 0.97) $\times$ (0.8 $\sim$ 8) & 512~$\times$~512~$\times$~(24~$\sim$~251) & - Largest abdominal dataset ($\uparrow\uparrow\uparrow$) \newline - Contains all liver masks (liver, tumors, vessels) & - N/A \\ \hline

CT-ORG \cite{CT-ORG} & \textcolor{black}{2019} & Liver (Tumors) \& Others & CT \, CE-CT PET-CT & \textcolor{black}{140 (119/21)} & (0.55 $\sim$ 1.37) $\times$ (0.55 $\sim$ 1.37) $\times$ (0.7 $\sim$ 5) & 512~$\times$~512~$\times$~(74~$\sim$~987) & - Builds up over LiTS and adds 9 more records \newline - Has other organs segmented & - Does not have liver vessels \\ \hline

\multirow{3}{*}{\centering CHAOS \cite{CHAOS}}  & \multirow{3}{*}{\centering\textcolor{black}{2019}} & Liver & CE-CT & \textcolor{black}{40 (20/20)} & (0.70 $\sim$ 0.80) $\times$ (0.70 $\sim$ 0.80) $\times$ (3 $\sim$ 3.2) & 512 $\times$ 512 $\times$ (77 $\sim$ 105) & \multirow{3}{3.8cm}{- Multiple modalities \newline - Large size ($\uparrow\uparrow$)} & \multirow{3}{2.8cm}{- Does not have liver tumors and vessels} \\ \cline{3-7}
    & & Liver, Kidneys \& Spleen & MRI & \textcolor{black}{120~(60/60)} & (1.36 $\sim$ 1.89) $\times$ (1.36 $\sim$ 1.89) $\times$ (5.5 $\sim$ 9) & 256 $\times$ 256 $\times$ (26 $\sim$ 50) & & \\ \hline
    
\end{tabu}
\end{table}
\end{landscape}

\section{Standard Segmentation Evaluation Metrics}\label{sec3}
In this section, we discuss the most considered metrics within the liver segmentation literature, highlighting the used notations, the inclusion/exclusion criteria, and the significance each metric presents. 

\subsection{Notations and Criteria}
\subsubsection{Notations}
Before discussing the metrics, we highlight the used notations.
\begin{itemize}
    \item $A$ refers to the ground-truth label voxels set
    \item $B$ is the predicted voxels set by the created models
    \item $|\cdot|$ is the set cardinality
    \item $||\cdot||$ represents the Euclidean distance
    \item $S(\cdot)$ indicates the set of surface voxels
    \item True positive ($TP$) is the set of correctly classified tissue of interest (TOI) pixels/voxels
    \item True negative ($TN$) is the set of truly classified background pixels/voxels, noting that background voxels describe any voxel which does not belong to the TOI of the study. 
    \item False positive ($FP$) is the set of incorrectly classified background pixels/voxels
    \item False negative ($FN$) is the set of incorrectly classified TOI pixels/voxels
\end{itemize}


\subsubsection{Inclusion/Exclusion Criteria}
All the metrics frequently used in the literature are included in this section highlighting the measurement each one conveys.

\subsection{Percentile Metrics}

\subsubsection{Jaccard Index (JI)}
JI is a fundamental metric to understand how close is the generated prediction in overlapping with the ground-truth label. It is also known as the Tanimoto index, or intersection-over-union (IoU) metric \cite{Heimann2009}. Equation \eqref{eq:JI} shows two equivalent definitions of the JI metric.

\begin{equation} \label{eq:JI}
    JI = \dfrac{|A \cap B|}{|A \cup B|} = \dfrac{TP}{TP + FP + FN}
\end{equation}

Intuitively, perfect prediction is when JI is equal to 1, meaning that $|A \cap B|$ is the same as $|A \cup B|$. In other words, there are no wrong predictions (i.e., $FP$ and $FN$ = 0), and the volumes are perfectly similar. In contrast, JI equating to 0 means that no intersection exists between the ground-truth and prediction, or $TP$ is 0, meaning that the TOI was completely misclassified.

\subsubsection{Precision/Positive Predictive Value (PPV)}
Precision aims to investigate the over-segmentation aspect of the model by dividing the total number of correctly classified TOI voxels over the total positively classified voxels (i.e., true and false) as indicated by equation \eqref{eq:Precision}.
\begin{equation} \label{eq:Precision}
    Precision/PPV = \dfrac{TP}{TP + FP}
\end{equation}

A value of 1 indicates an ideal segmentation scenario for correctly classifying background voxels. In contrast, a value of 0 is the extreme case of incorrectly classifying all TOI voxels.

\subsubsection{Recall/Sensitivity/True Positive Rate (TPR)}
Recall, on the other hand, investigates the under-segmentation aspect of the model, by dividing the correctly classified TOI voxels over the ``actual'' number of TOI voxels, as shown by equation \eqref{eq:Recall}.
\begin{equation} \label{eq:Recall}
    Recall/Sensitivity/TPR = \dfrac{TP}{TP + FN}
\end{equation}

Recall varies between 0 and 1, where 1 indicates perfect segmentation of all TOI voxels, and 0 indicates the exact opposite.

\subsubsection{Dice Similarity Coefficient (DSC)}
DSC (or Dice) is the F1 Score counterpart for images, which is a harmonic mean of both precision and recall. In a sense, it measures the similarity between ground-truth set $A$ and generated prediction $B$. The original DSC for a single image is defined in equation \eqref{eq:DSC}.

\begin{equation} \label{eq:DSC}
    DSC = 2\dfrac{|A \cap B|}{|A| + |B|} = \dfrac{2TP}{2TP + FP + FN}
\end{equation}

Similar to the JI metric, the two extreme cases are 0 and 1, where the former emphasizes the absence of any similarity and the latter shows the perfect similarity between $A$ and $B$.

The organizers of the LiTS workshop further formulated two metrics from DSC, highlighting a key difference, how the DSC of each case is summed. By this distinction, Dice/case is aimed to account equally for both small tumors and large tumors, and not be highly influenced by the large ones \cite{Patrick2019}.\\

\textbf{\textit{Global DSC (GDSC)}} encapsulates the segmentations of all volumes and compares them to all respective labels in a single shot as if all the volumes were concatenated into one. Thus, having a similar effect to equation \eqref{eq:DSC} for the whole test set.\\

\textbf{\textit{Dice/Case (DPC)}} calculates DSC per volume and then averages the DSC of all the volumes in the test set. Adjustment to the DSC formula is shown in equation \eqref{eq:DPC}.

\begin{equation} \label{eq:DPC}
    DPC = \dfrac{1}{N}\mathlarger{\sum_{i=1}^{N}}2\dfrac{|A_i \cap B_i|}{|A_i| + |B_i|} = \dfrac{1}{N}\mathlarger{\sum_{i=1}^{N}}\dfrac{2TP_i}{2TP_i + FP_i + FN_i}
\end{equation}
where $N$ represents the number of volumes in the testing set and $i$ represents the $i^{th}$ volume from the $N$ volumes.

\subsubsection{Specificity/True Negative Rate (TNR)}
As depicted in equation \eqref{eq:Specificity}, specificity investigates the model's capability in classifying background voxels correctly.
\begin{equation} \label{eq:Specificity}
    Specificity/TNR = \dfrac{TN}{TN + FP}
\end{equation}

Ranging between 0 and 1, the former denotes a misclassification of all background voxels, and the latter resembles a proper classification of all background voxels.

\subsubsection{False Positive Rate (FPR)/Fallout}
As shown by equation \eqref{eq:FPR}, and complementary to the specificity definition, it highlights the amount of error the model is making when classifying background voxels.
\begin{equation} \label{eq:FPR}
    FPR/Fallout = 1 - Specificity = \dfrac{FP}{FP + TN}
\end{equation}

Contrary to specificity, a value of 0 is a good indicator of the model's ability in predicting background voxels. On the other hand, a value of 1 is an extreme scenario where the model wrongly classified all background voxels.

\subsubsection{Receiver Operating Characteristic (ROC) and Area Under ROC Curve (AUC)}
\begin{figure}
    \centering
    \includegraphics[trim = {0.3in 0.3in 0.3in 0.3in},clip, width=0.75\linewidth]{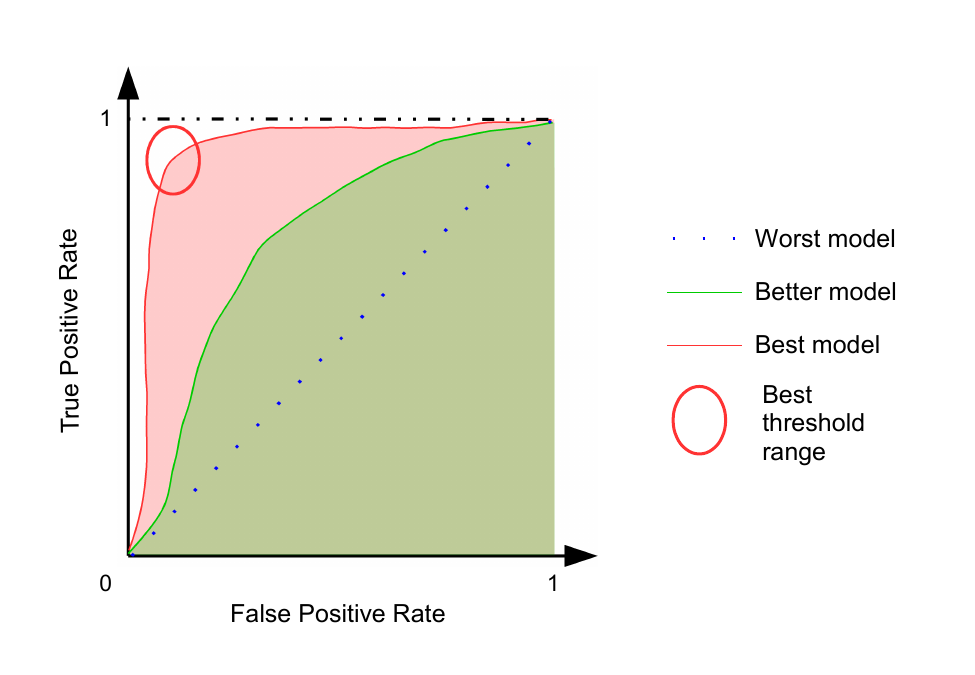}
    \caption{ROC and AUC for multiple models}
    \label{fig:roc_auc}
\end{figure}

ROC curve takes advantage of the TPR and FPR metrics to measure the contribution of using a certain threshold, for which, the voxel will be classified as either a TOI or background voxels. It encompasses the resulting TPR and FPR for every threshold between 0 and 1, showing the ideal threshold range to be at the top left corner for a specific model. Multiple models with varying thresholds can be compared in between by the AUC metric, where the larger the area is, the better the model is, for different threshold values. Fig.~\ref{fig:roc_auc} illustrates the underlying benefits of using ROC curves in comparing trained models.

\subsubsection{Volumetric Overlap Error (VOE)}
VOE is the complementary metric of JI, which is known as Jaccard distance, knowing that VOE is a special case for volumetric sets. It measures the spatial error represented between the voxels of $A$ and $B$ \cite{Heimann2009} and is described by equation \eqref{eq:VOE}.

\begin{equation} \label{eq:VOE}
    VOE = 1 - \dfrac{|A\cap B|}{|A \cup B|} = \dfrac{FP + FN}{TP + FP + FN}
\end{equation}

VOE ranges between 0 and 1, where the former means that the voxels of $B$ are perfectly and correctly lying over $A$'s voxels, and the latter indicates the absence of overlapping voxels between the voxels of $A$ and $B$.

\subsubsection{Relative Volumetric Distance (RVD)}
RVD measures the difference between volume $A$ and $B$, and is an indicator of whether the set of voxels encompassed by $B$ is an under- or over-segmentation by comparing it with $A$'s voxels \cite{Heimann2009}. Equation \eqref{eq:RVD} highlights this metric.

\begin{equation} \label{eq:RVD}
    RVD(A,B) = \dfrac{|B| - |A|}{|A|}
\end{equation}

This metric can be positive, negative or zero, whereas being positive indicates that $B$ is over-segmenting the original volume, being negative indicates an under-segmentation case, and being zero as having identical volumes. RVD should not be used alone as it does not necessarily indicate an overlap between $A$ and $B$ \cite{Heimann2009}.

\subsection{Distance Measurements}
The distance measurements extensively used in literature are mentioned here, each measurement captures a certain spatial aspect, and all of them are measured in mm.

\subsubsection{Average Symmetric Surface Distance (ASD)}
\begin{figure}
    \centering
    \includegraphics[trim = {0.75in 1in 0.8in 1in},clip, width=0.7\linewidth]{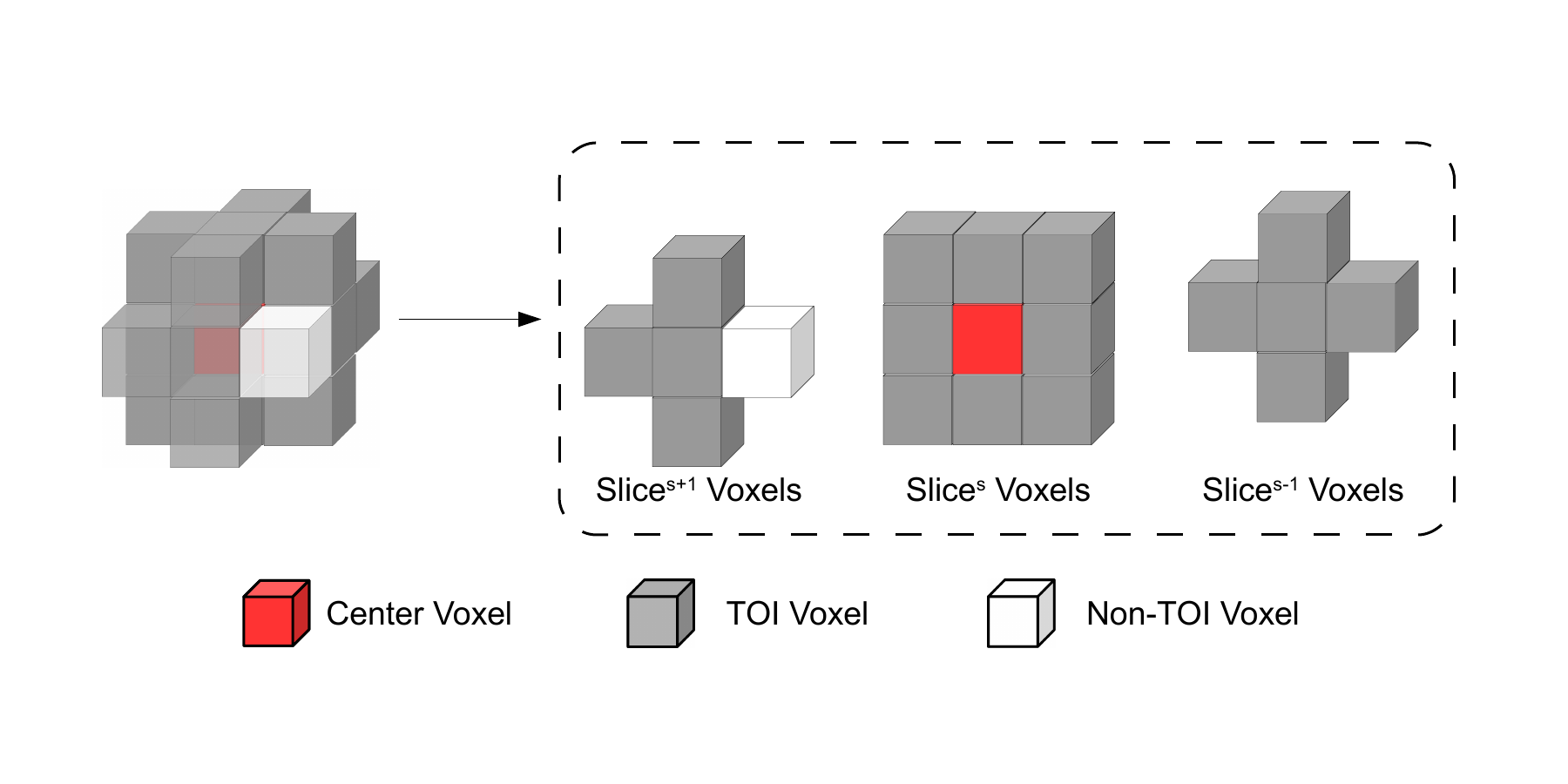}
    \caption{A surface voxel (red) with 1 non-TOI adjacent voxel (white) and 17 TOI voxels (grey)}
    \label{fig:surface_voxel}
\end{figure}

ASD measures the minimum distance that can be found between a surface voxel in $A$ to another surface voxel in $B$. Since it is a symmetric metric, the same applies to $B$ with respect to $A$. Then, the average is taken over all the calculated distances. Surface voxel is a name given to a voxel with at least one non-TOI voxel (i.e., background voxel) from its 18-neighboring voxels, as shown in Fig.~\ref{fig:surface_voxel}. To define ASD, we first have to define the minimum distance between an arbitrary voxel $v$ and $S(A)$:

\begin{equation} \label{eq:min_distance}
    d(v,S(A)) = \min_{s_A \in S(A)}||v-s_A||
\end{equation}
where $s_A$ is a single surface voxel distance from the surface voxels set $S(A)$.

Using equation \eqref{eq:min_distance}, we can now define ASD as following:
\begin{equation} \label{eq:ASD}
    ASD(A,B) = \dfrac{1}{|S(A)|+|S(B)|}\left(\sum_{s_A \in S(A)} d(s_A, S(B)) + \sum_{s_B \in S(B)} d(s_B, S(A)) \right)
\end{equation}
From equation \eqref{eq:ASD} and the definition of Euclidean distance, it can be seen that this metric is always positive. The value converges to 0 when the highest spatial similarity is achieved. However, the larger the value, the worse the overlap between volumes $A$ and $B$ is noticed, and dissimilarity starts to be observed.

\subsubsection{Root-Mean-Square Symmetric Surface Distance (RMSD)}
RMSD is significantly related to the ASD definition described by equation \eqref{eq:ASD}. The root-mean-square of ASD is the summation of distances squared under the square root as defined by equation \eqref{eq:RMSD}.
\begin{equation} \label{eq:RMSD}
    RMSD(A,B) = \sqrt{\dfrac{1}{|S(A)|+|S(B)|}\left(\sum_{s_A \in S(A)}d^2(s_A, S(B)) + \sum_{s_B \in S(B)}d^2(s_B, S(A))\right)}
\end{equation}

The benefit gained by defining such a metric is the weight bestowed over large deviations, making the metric more sensitive to outliers \cite{Heimann2009}. Equivalently, the best value for this metric should be 0, and the bigger it is, the worse the volumes' overlap is. 

\subsubsection{Maximum Symmetric Surface Distance (MSD)/Hausdorff Distance (HD)}
MSD, famously known as HD as well, searches for the maximum distance, defined by equation \eqref{eq:min_distance}, that can be found between volumes $A$ and $B$. 
\begin{equation} \label{eq:MSD}
    MSD(A,B) = \max\left\{\max_{s_A \in S(A)}d(s_A, S(B)), \max_{s_B \in S(B)}d(s_B, S(A) )\right\}
\end{equation}

This metric gives the maximum distance error between $A$ and $B$, and thus, is extremely sensitive to outliers \cite{Heimann2009}.


\subsection{\textcolor{black}{Discussion}}
\textcolor{black}{
Selecting the appropriate metric for a specific medical image segmentation task is of utmost importance. The abovementioned evaluation metrics are essential for a transparent, objective, and fair performance comparison and assessment; however, little concern was given to discussing their limitations. Overall, one main advantage of the percentile metrics over the distance-based measures is their fixed value range ([0; 1], also reported as [0; 100]). This helps easily compare the results reported in different studies, unlike the distance-based metrics with their unfixed value range ([0; $\infty$)). However, the main problem with percentile-based metrics is that they account only for the number of correctly or miss-classified pixels without reflecting their spatial distribution.
On the other hand, another issue with distance-based metrics is the variety of methods used to define region borders based on the selected neighborhood size \cite{taha2015metrics}.
Additionally, as reported in \cite{Voiculescu2015AnOO}, using only one measure among the abovementioned ones to evaluate automatic liver segmentation techniques is unsuitable to serve as a reliable metric that can reflect all the aspects of segmentation accuracy and errors. However, adopting multiple metrics can avoid their limitations and make them complement each other.}

\section{Categorization of ML Techniques} \label{sec4}
In this section, surveyed works are categorized based on how the volumes are input to the ML algorithms, followed by a detailing of the ML algorithms.

\subsection{Input shape}
\begin{figure}[!ht]
\centering
\includegraphics[width=\linewidth, trim={0.7in 0in 0.7in 0in}, clip]{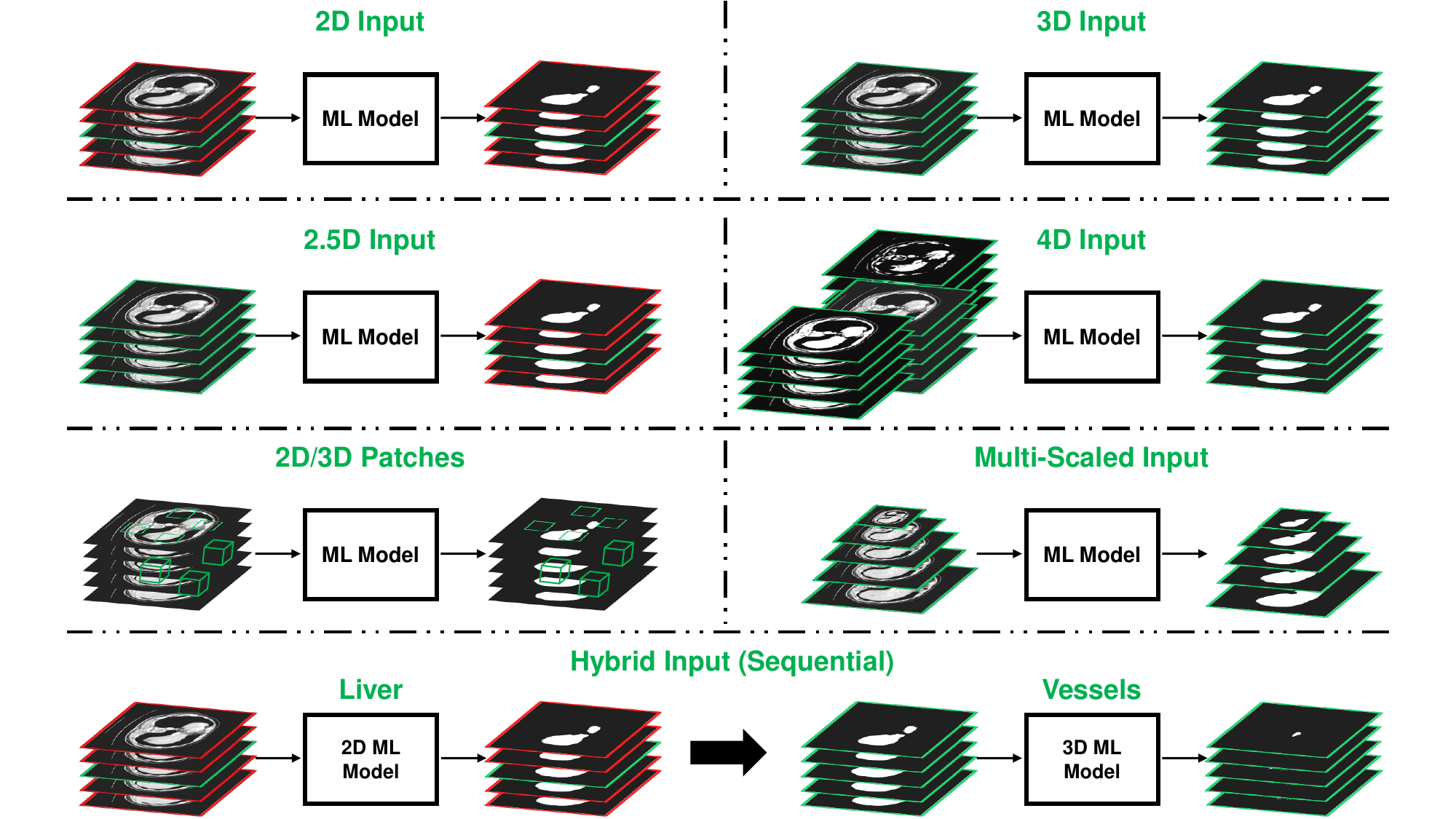}
\caption{\textcolor{black}{Different input shapes inserted into an ML model, where green indicates slices being inserted/outputted to/from an ML model and red means that the slices are not involved in that specific segmentation step for the green slice (CT images from \cite{MSDC}).}}
\label{input_shapes}
\end{figure}

When reviewing the literature, slices are inserted into the networks in different shapes and dimensions. CNNs, by their many forms, can accept inputs with different dimensions. The importance of discussing the input dimensions prevails when we know that these dimensions also affect the CNN architecture, where filter sizes, convolutional layers, and pooling layers will be designed differently. This subsection highlights works utilizing different input dimensions: 2D, 3D, 2.5D, 4D, hybrid, patches, or multi-level scaled-down slices. \textcolor{black}{Fig.~\ref{input_shapes} visually explains the difference between different input shapes while being fed into an agnostic ML algorithm.}

\subsubsection{2D Input}
Originally, CNNs are developed to work with inputs in their 2D form, i.e., images. Many works opt for the utilization of 2D inputs, where slices from the imaging modalities are inserted into the CNN in a slice-by-slice fashion. Each slice is segmented in a single forward pass isolated from adjacent slices as in \cite{Christ2016} for example, where a 3-dimensional conditional random field (3D CRF) has to be used to impose 3D context. If the slices were fed sequentially (for the same volume) into the CNN model but randomizing the volumes, the model could implicitly understand that there is a 3D context within. However, the emphasis is not strong, and this 3D context quickly dissipates if the training was randomized within volumes as well.

\subsubsection{3D Input}
3D inputs are used to involve the volumetric context in the segmentation task. It is helpful to use slices in their 3D form, meaning multiple slices are inserted into a single CNN, and the segmentation is carried out in one shot over all of them in the forward pass. Consequently, the network learns the importance of 3D context around the TOI; thus, inter-slice information is preserved compared to the 2D counterpart. For instance, the famous 2D U-Net \cite{Ronneberger2015} and its 3D counterpart \cite{Cicek2016} demonstrate the changes in the network architecture that follow the transition from receiving a 3D input instead. However, complications are imposed when using the 3D volumes because of the limitations in the GPU memory and the accompanying heavy calculations required by the implementation of 3D CNNs. Thus, researchers would then have to deal with patches of the original volume or a coarsely down-scaled version of it.

\begin{figure}[!ht]
\centering
\includegraphics[width=\linewidth]{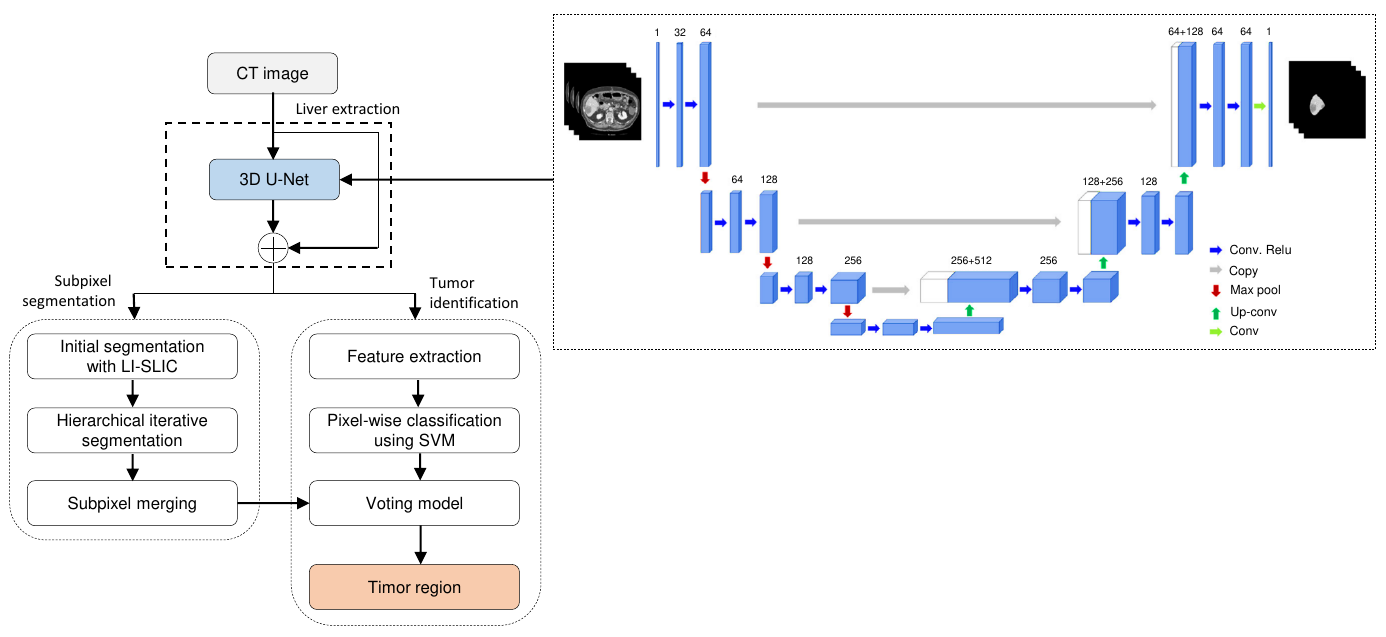}
\caption{\textcolor{black}{Flowchart of the 3D U-Net based liver segmentation scheme proposed in \cite{di2022automatic}.}}
\label{3D_U-Net_flowchart}
\end{figure}

\textcolor{black}{To reflect the differences between outside-liver, tumor, and inside-liver tissues, liver and tumor segmentation from 3D abdominal CT volumes are performed in \cite{chi2021x} using a multi-branch U-Net-like model, namely X-Net. Typically, to better extract intra-slice features of liver and tumors, a pyramid-like convolution structure for inner-liver feature extraction and an up-sampling branch for liver region recognition are embedded in the back-bone Dense-UNet. Moreover, conventional 3D U-Net is simplified by adopting fixed-size convolutional kernels (3$\times$3 in the x-y plane) and applying it as a 3D counterpart to aggregate contextual information along the z-axis from the stacked, filtered CT slices. This helps inhibit the influence of neighboring pixels and greatly alleviates the computational burden. In \cite{di2022automatic}, the authors accurately segment liver tumors from CT images by using 3D U-Net to detect liver regions and alleviate the computational cost of the segmentation. Typically, liver regions are first extracted using a 3D U-Net before dividing them into homogeneous superpixels by applying a hierarchical iterative segmentation strategy, which relies on local-information-based simple linear iterative clustering (LI-SLIC). Consequently, this enables classifying every pixel in the liver regions into non-tumor or tumor using SVM with its texture features and local intensity. 
To segment the liver and the tumor, the authors in \cite{alalwan2021efficient} develop a 3D semantic segmentation deep learning (DL) model, namely 3D DenseU-Net-569. The latter is a fully 3D semantic segmentation model, which encompasses lower training parameters and a considerably deeper network. It also relies on depthwise separable convolution (DS-Conv) instead of conventional convolution. Fig.~\ref{3D_U-Net_flowchart} presents the flowchart of the 3D U-Net-based liver segmentation scheme proposed in \cite{di2022automatic}.}

\subsubsection{2.5D Input}
2.5D is the middle ground between the 2D and 3D inputs, where it utilizes the 3D context and information while restraining the segmentation to a single slice in a single forward pass. The input is a slice in the middle of its neighboring slices, creating an odd number of slices $2(k-1)$ inserted into the CNN model. In essence, the CNN is built to receive inputs of 2D nature with multiple channels; however, the key difference lies in its output, where the segmentation mask is generated for the center slice only, and the neighboring slices serve only as context and 3D spatial information providers for the model. The idea in itself is not new, it is clearly stated in \cite{Roth2014new}, but many works opt for this method as it harnesses the benefits of both inputs' dimensions and disposes of their disadvantages \cite{Han2017,Li2018,Chlebus2018, Vorontsov2018,Wang2019pairwise,Zheng2019,He2020}, and others. \\

\textcolor{black}{
Because 2D DL-based segmentation models are less accurate while 3D ones are accurate but large and computationally expensive, Tian et al. \cite{tian2021fully} automatically annotate functional regions of the liver using a 2.5D class-aware DNN with spatial adaptation. This framework is based on analyzing abdominal images using a ResU-Net model, which (i) adequately selects a pile of adjacent CT slices as input, generates the center slice, and (iii) automatically annotates the liver functional regions. Besides, in \cite{zhang2021liver}, 2.5D UV-Net with multi-scale convolution is utilized to segment liver tumors. In doing so, multi-scale feature extraction is performed with a similar computational cost to 2.5D UV-Net. This enables mining structured data, reducing data redundancy, strengthening independent characteristics, making features sparse, and enhancing network efficiency and capacity. In \cite{lv20222}, automatic liver and tumor segmentation are conducted using a lightweight Inception convolution architecture with residual connections, which significantly enables reducing the model's parameters. Accordingly, a 2.5D lightweight RIU-Net model is adopted, which deploys DSC and binary cross entropy (BCE) loss to reach fast convergence and low fluctuations in training. Similarly, in \cite{han2021boundary}, a cascaded 2.5D FCN model is considered for liver and tumor segmentation from 3D medical images. Additionally, the FCN model is augmented with a boundary loss incorporating boundary information, area, and distance, to learn more contour features and boundaries from the 3D images. In \cite{wardhana2021toward}, a modified 2.5D SegNet model is used for liver and tumor segmentation by: (i) utilizing the long-range connection from U-Net; and (ii) implementing the short/skip connection that is generally found in ResNet.}


\subsubsection{4D Input}
As controversial as it sounds, the 4D concept comes from the MRI modality since it generates multi-phase 3D volumes of the same shape with a difference in temporal acquisition. They can be grouped when used for segmentation. In \cite{Ivashchenko2020}, the multi-phase volumes are used in 4D K-means clustering aided by active contour refinement. On the other hand, in \cite{Takenaga2019}, the 3D volumes are inserted into a 3D ResNet-based CNN, where the input is multi-channel (effectively making it 4D). However, this does not reflect architecture change as it can be resolved by channel depth design at the first layer.

\textcolor{black}{In \cite{zheng2022automatic}, a 4D DL network built upon 3D convolution and convolutional LSTM (C-LSTM) for HCC lesion segmentation. This DL module uses 4D data corresponding to dynamic contrast-enhanced (DCE) MRI images to assist liver tumor segmentation. Accordingly, 3D spatial domain features from every DCE phase are extracted using a shallow 3D U-net model before applying a 4-layer C-LSTM for time domain information exploitation.}

\begin{figure}[!ht]
\centering
\includegraphics[width=\linewidth]{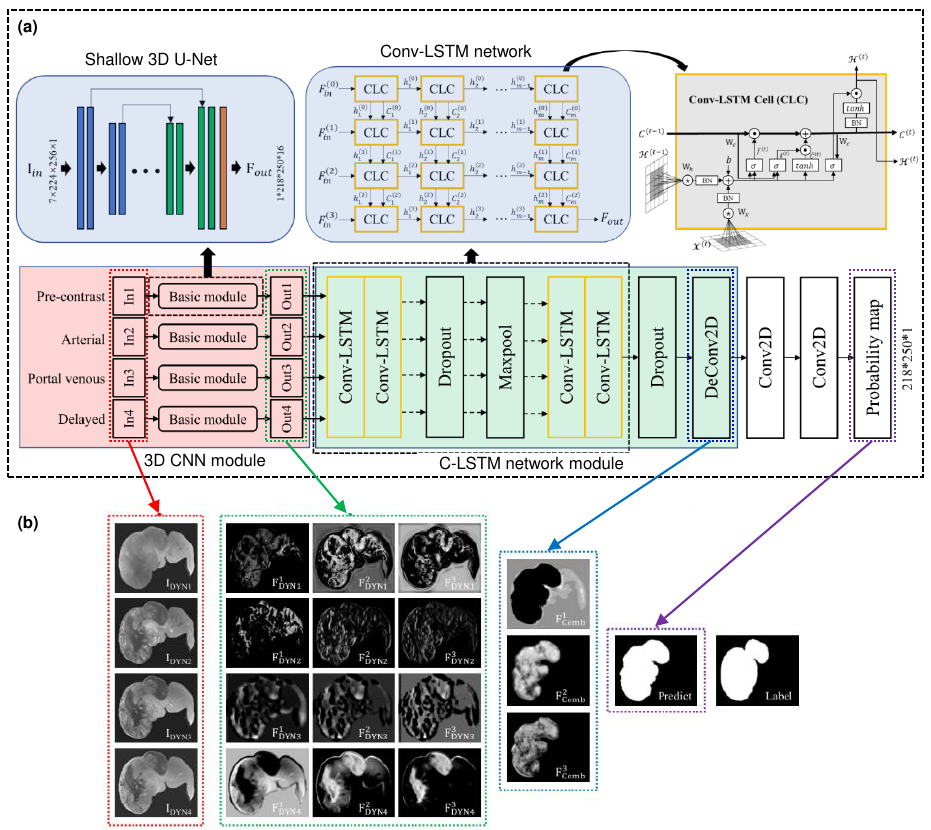}
\caption{\textcolor{black}{Overall 4D DL framework for HCC segmentation proposed in \cite{zheng2022automatic}: (a) the architecture of the 4D model, and (b) the feature map analysis for the case of large tumors with internal inhomogeneity.}}
\label{4D_flowchart}
\end{figure}

\textcolor{black}{Fig.~\ref{4D_flowchart} presents the overall architecture of the 4D framework for HCC segmentation. This includes (i) a 3D CNN module (in pink), and (ii) a C-LSTM network module (in green). Additionally, a shallow 3D U-Net has been utilized to extract spatial domain information in the pre-contrast, farterial, portal venous, and delayed phases, separately. Moreover, a 4-layer Conv-LSTM network has been developed to exploit time domain information via multiple DCE phases ($m$ refers to the number of layers of the C-LSTM network, and in this case, $m=4$).}

\subsubsection{2D \& 3D Patches}
Taking patches means that the programmer extracts small portions of the slice/volume and then inputs them to the fully convolutional network (FCN) to perform the segmentation. Theoretical background for why it works is provided in \cite{Chen2015}, and some of the reasons are mentioned here. Especially for the 3D case, GPU and extensive computations are the barriers to evaluating volumes in one shot. Thus, researchers opt to take 3D chunks from the volumetric scan to process, which can have homogeneous dimensions as in \cite{Wang2019abdominal,Zhang2018liver, Cheema2019}, or can have heterogeneous dimensions \cite{Roth2018article}. It is worth noting that in \cite{Roth2018article}, the effect of overlapping patches is studied against non-overlapping ones. In general, for both 2D and 3D, it helps the model to generalize better to unseen real-life scenarios when the segmentation is done over patches instead of the whole slice/volume. Moreover, as in \cite{Li2015}, 2D patches are very convenient when the TOI is small, e.g., tumors and vessels within the liver, as processing the whole liver would be redundant when segmenting such tissues.

\subsubsection{Multi-Scaled Input}
In more recent works, researchers opt to fuse segmentations on different scales. Such algorithms take multi-scaled inputs, concatenate, and fuse them on many levels of the network to generate the segmentation mask either in a sequential manner as in \cite{Christ2016} or in one shot as in \cite{Fang2020deep}.

\subsubsection{Hybrid Input (Sequential)}
Hybrid input emphasizes the employment of sequential models with different inputs' dimensions, whereas the model can utilize an input with distinct dimensions compared to the previous/following one. To increase segmentation accuracy, \cite{Jin2018} utilized a 2D network to acquire a coarse liver segmentation, which is sequentially inserted into a 3D network for segmentation refinement. Another approach is jointly using 2D and 3D networks for liver and tumor segmentation and fusing both networks' outputs \cite{Li2018}. Other methodology uses the first 2D CNN to segment the liver and large tumors and the sequential 3D CNN to focus on segmenting small ones \cite{Dey2020}. It is worth noting that the training methodology followed in \cite{Liu2018} initially relied on developing the weights in a 2D network, which is then extrapolated into their 3D counterpart. 

\subsection{Categorization based on Classified Tissues}
Since it is desired to obtain accurate and real-time results for the liver delineation problem automatically, it is intuitive that ML algorithms are utilized. However, providing a measure for selecting a particular ML algorithm for a specific segmentation scenario would certainly help, especially with the technological advancements we are currently witnessing. Thus, in this subsection, the determination of the ML algorithms, unsupervised and/or supervised, is shown based on the application area, liver parenchyma, tumors, and/or vessels within the liver.

\subsubsection{Liver}
\begin{figure}[!ht]
    \centering
    \includegraphics[width=\linewidth]{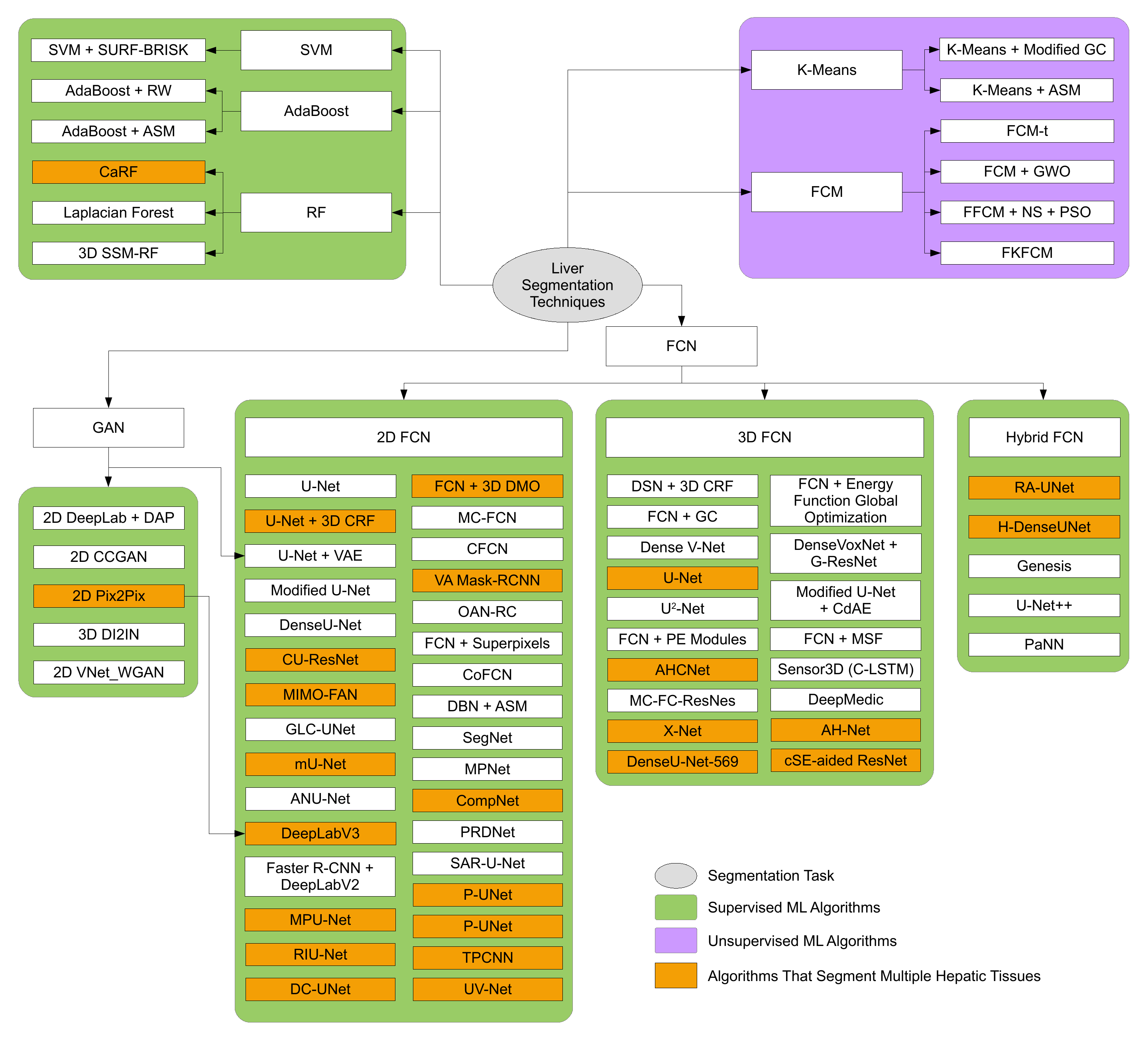}
    \caption{Liver segmentation techniques from the reviewed literature}
    \label{fig:liver_map}
\end{figure}

\textcolor{black}{The main tissue/organ that is considered in this survey is the liver. Thus, it naturally it is the first tissue that we need to to conduct the survey on, since it encapsulates all the tissues residing within. Fig.~\ref{fig:liver_map} describes the basis of different algorithms used for the liver segmentation challenge.}

\vspace{1em}
\textbf{\textit{Miscellaneous Unsupervised}}

Unsupervised ML algorithms have a fair share in the liver segmentation task. The reason for their efficacy is that the liver is a single large continuous organ with relatively similar pixels' intensities in the same CT/MRI volume. The most prominent unsupervised algorithms are the k-means clustering and fuzzy c-means (FCM) clustering, where the former only allows the belonging of a certain pixel to a specific centroid (hard), while the latter allows for the pixel to belong to multiple centroids, with a certain value (soft). In \cite{Huang2018}, k-means clustering is used for liver localization in CT slices as a basis for thresholding, followed by modified GC segmentation. However, in \cite{Ivashchenko2020}, a 4D k-means is utilized on multi-phase MRI volumes for liver segmentation aided by the active shape modeling technique. It is worth noting that vessel extraction is implemented through multi-scale vesselness filters.

On the other hand, FCM clustering in \cite{Alahmer2016} calculates the degree of belonging for each pixel to three cluster classes, where one of these centroids represents liver pixels. In \cite{Ali2014} the study focuses on choosing an optimum threshold, FCM-t, which best determines the degree of belonging a pixel should convey to be considered a liver pixel. Other researchers use FCM, or an enhanced version of it, for the liver delineation task. For instance, FCM is used with the grey wolf optimization (GWO) algorithm in \cite{Sayed2016}, while a fast FCM (FFCM) is utilized with neutrosophic sets (NS) and particle swarm optimization (PSO) in \cite{Anter2018}. Lastly, in \cite{Cai2019}, a fast kernelized FCM (FKFCM) is used to segment the liver.\\

\noindent\textbf{\textit{Miscellaneous Supervised}}

Laplacian forest (LF), an improved version of random forests (RF), is used in \cite{Lombaert2014} for the liver segmentation task among other organs. Other works use RF as a landmark detection tool for 3D SSM \cite{Norajitra2017}, while cascaded random forest (CaRF) classifiers are used for liver parenchyma segmentation in \cite{Treilhard2017}.

In an ensemble of weak classifiers fashion, adaptive boosting (AdaBoost) is used to segment the liver with the aid of random walks (RW) in \cite{Zhang2015}. A similar work uses the same combination with extra improvements on the RW algorithm in \cite{Zheng2017automatic}, and finally, a three-level ASM is guided by the AdaBoost algorithm in \cite{He2016}.

In \cite{Liang2018}, speeded up, robust features (SURF) (blob-type), and binary, robust, invariant, scalable key points (BRISK) (corner-type) features are used in a top-bottom flow, aided by the support vector machine (SVM) in the bottom-up counterpart pathway to segment the liver.\\

\noindent\textbf{\textit{GAN}} 

A generative adversarial network (GAN) is also employed for this important task, where in \cite{Yang2017}, a 3D deep image-to-image network (DI2IN) is used to segment the liver. On the other hand, in \cite{Chen2019}, cascaded conditional GANs (CCGAN) are used for the same task but in 2D form. In \cite{Zheng2019}, a GAN network incorporating a deep atlas prior (DAP), where the generator, based on DeepLab (ResNet101), is used for the liver segmentation, and the discriminator is a simple 2D FCN used to challenge the generator segmentation.\\

\noindent\textbf{\textit{2D FCN}} 

FCNs have become the natural choice from various ML and artificial intelligence (AI) algorithms when the data come in more complex forms than the 1D form. FCNs are a sector of CNN algorithms, where fully connected layers at the end of the network are replaced by convolutional ones, reducing the number of parameters. \textcolor{black}{Goodfellow et al. mention in \cite{goodfellow2016deep}: \enquote{The hierarchy of concepts allows the computer to learn complicated concepts by building them out of simpler ones. If we draw a graph showing how these concepts are built on top of each other, the graph is deep, with many layers. For this reason, we call this approach AI deep learning.} This statement means that a CNN is considered deep if layers build complex structures that are based on simpler structures outputted from the previous layer}. Moreover, it is more appropriate for the segmentation task as the output's form of such networks is similar to the input. Nonetheless, FCNs are used for various problems and purposes. For instance, in \cite{Tian2018}, a 2D FCN is utilized for liver segmentation, which is then used for diagnosis report generation. In \cite{Wang2018}, a 2D FCN-8s training is done via a newly-devised sample selection idea named relaxed upper confident bound (RUCB). In \cite{Yuan2017}, cascaded 2D FCN (CFCN) is used for liver segmentation, where the first FCN coarsely segments the liver and the second one refines it. In \cite{Zheng2017}, another 2D FCN is used for the liver segmentation, followed by a 3D deformable model optimization (3D DMO) based on local cumulative spectral histograms and non-negative matrix factorization (NMF).
In \cite{Jansen2019}, a 2D multi-channel FCN (MC-FCN) takes six slices as input from multi-phase MRI imagery, where the used structure outperforms the U-Net on the utilized dataset. In \cite{Qin2018}, superpixels are computed, forming a map using simple linear iterative clustering (SLIC) algorithm, and then the map is introduced into 2D FCN to segment the liver. In \cite{aghamohammadi2021tpcnn}, a two-path CNN (TPCNN) is used to segment the liver using patches of varying sizes, with a novel encoding approach to extract features from CT images. The input is the image itself, the Z-score normalized one, and the encoded image using the local direction of gradient (LDOG) algorithm. 

An important FCN architecture that revolutionized the biomedical segmentation field is the U-Net \cite{Ronneberger2015}, playing a similar role to the AlexNet, but for the biomedical field. Thus, it was natural for some researchers to use it. In \cite{Christ2016,Christ2017}, a 2D FCN following the U-net architecture is utilized along with a 3D CRF for liver segmentation. In \cite{Ouhmich2019}, a 2D U-Net is used as the main model, while SegNet is utilized in \cite{Nanda2019}. In \cite{Zhang2020a}, the U-Net acts as a coarse liver segmenter; however, in \cite{Mendizabal2020}, U-Net is used as a replacement for the finite element method (FEM) to approximate the elastic deformation caused in hyperelastic objects, such as the liver. Interestingly in \cite{Wang2020}, the 2D U-Net is used to segment the liver, but the work does not focus on the model's accuracy because the target is to examine whether a slice can be used to make a diagnostic decision or not. In \cite{Li2021}, U-Net is used as the main model, but components such as a Bi-ConvLSTM are integrated to enhance the liver's edge capturing. 

Other researchers are inspired by the U-Net structure. A 2D FCN (modified U-Net version) is employed for segmenting the liver parenchyma, excluding vessel ducts, from T1-MRI scans \cite{Irving2017}. In \cite{Liu2020}, 16 phases (echoes) of the same slice are generated by  employing the multi-echo gradient from the MRI imaging modality, and the kernels at the first layer of the 2D U-Net are modified to accept 16 slices as an input. In a similar approach, unenhanced multi-echo spoiled gradient-echo slices from MRI scans are initially used to train a 2D U-Net, followed by a transfer learning (TL) training step on CE-CT and CE-MRI to segment liver from both modalities \cite{Wang2019automatedCT}. Additionally, in \cite{Vorontsov2018}, an ensemble of three U-Net-like 2D FCN models is used for the liver segmentation task, and the final mask is the average of those three. \cite{Maaref2020} utilized the same segmentation network as in \cite{Vorontsov2018}, but with more interest in tumor classification. In \cite{Seo2020}, the skip connections between the encoder and decoder are modified to eliminate the redundant inclusion of low-resolution information, and the network is named modified U-Net (mU-Net). In \cite{Guo2020}, the semantic segmentation of multiple organs is carried via a 2D ResNet equipped with partially dilated convolutions and multiple concatenations and fusion stages.

On the one hand, a multi-planar network (MPNet) is employed to segment the liver in any view (transversal, sagittal, or coronal) \cite{Wang2019automatic,Chen2019a}. In their work, an ensemble of three MPNets is trained to segment the liver from each view, and in the end, the segmentation mask from the three MPNets are fused to generate the final output. \textcolor{black}{A similar approach has been utilized in the organ-attention networks with reverse connections (OAN-RCs) developed by \cite{Wang2019abdominal}, where reverse connections are constructed to pass semantic information to the lower layers for coarse organ segmentation, followed by a fine-tuning stage. Finally, the output of the 3 OAN-RCs is combined through statistical similarity fusion.} On the other hand, in \cite{Perslev2019}, a multi-planar U-Net (MPU-Net) is utilized to capture the organ of interest from different viewing angles (generalizing to more views than the three conventional ones) and, similarly, fusing the output of all planar segmentation to generate the final output. In \cite{Yang2019}, a domain adaptation (DA) pipeline is created because the authors aim to create an algorithm that achieves great results on both CT and MRI scans. The first module is concerned with finding a common space between CT and MRI via variational autoencoders (VAE) and GANs. The second module takes the common space output from the first module and inserts it into a 2D U-Net to segment the liver, outperforming a CycleGAN-based solution. In \cite{Cheema2019}, a 2D liver extraction residual convolutional network (LER-CN), similar to U-Net architecture, is utilized to segment the liver from low-dose CT scans using two main components: noise removal component (NRC) and structural preservation component (SPC). More modifications have been applied to the U-Net structure. For instance, in \cite{Zhang2018}, a 2D FCN based on U-Net is equipped with ResNet dense forward connections (U-ResNet) for liver segmentation in digitally reconstructed radiographs (DRR) from X-rays via a task-driven generative adversarial network (TD-GAN). In \cite{Xi2020}, cascaded U-ResNet (CU-ResNet) is used for liver segmentation, concatenating the middle outputs from the liver U-ResNet with the corresponding output layers in the lesions' network. The work also aims to compare different loss functions, creating an ensemble of models incorporating the different loss functions \cite{Xi2020}. In \cite{Han2017}, densely-connected U-Net (DenseU-Net) is used for the liver segmentation task, and in \cite{Ahn2019}, a comparison between FusionNet and atlas-based segmentation models is conducted, proving the efficacy of the former to be used in a future clinical environment. A similar architecture is utilized in \cite{He2020}, where the 2D FCN is based on DenseU-Net, but interestingly, utilized a shallower decoder scheme and did not witness any reduction in the segmentation performance for the liver and other organs. In \cite{Tian2019}, both global and local context U-Net (GLC-UNet) are used to incorporate the global and local context, which also attempts to create Couinaud segmentation of the liver. In another study, a multiple-input and multiple-output feature abstraction network (MIMO-FAN) model adapted the U-Net architecture to generate multi-scale outputs for multi-scale inputs and fusing them to achieve the final output for liver in \cite{Fang2020deep}, and on partially labeled datasets for multiple organs in \cite{Fang2020multi}.

Sometimes a different backbone architecture is preferred by some of the researchers. For instance, a 2D FCN based on volume attention Mask-RCNN (VA Mask-RCNN) to incorporate volume information is employed for liver segmentation \cite{Wang2019volumetric}. In \cite{Xia2019}, a 2D FCN based on DeepLabV3 is used for liver segmentation, followed by Pix2Pix GAN in a two-player game competition to enhance the segmentation mask. In \cite{Tang2020}, Faster R-CNN is used for liver localization, while a DeepLabV2 network is used for the segmentation. In \cite{Dey2020}, a complementary network (CompNet) is employed for the segmentation task by attempting to incorporate non-TOI pixels into the learning of TOIs ones \cite{Dey2018}. A pairwise segmentation technique for sharing supervised segmentation between two paths is investigated by the conjugate FCN (CoFCN) \cite{Wang2019pairwise}, where it takes 2.5D input and learns from adjacent slices explicitly what the segmentation mask should be. In \cite{Ahmad2019}, 2D deep belief network (DBN) is deployed to segment the liver, aided by ASM for post-processing refinement.

\textcolor{black}{
To overcome the lack of interlayer information in 2D CNN models that can cause profound loss of segmentation performance, the authors in \cite{ma2021liver} develop a 2.5-D VNet\_WGAN.
Moving on, to avoid the sensitivity of liver segmentation models to heterogeneous pathologies and fuzzy boundaries, mainly when the data is scarce, 3D CNN and a hybrid loss function are deployed in \cite{tan2021automatic}. Typically, compressed codes of liver shapes are obtained using an autoencoder before training a liver segmentation network with a hybrid loss function. 
In \cite{meng2021two}, liver and tumor segmentation is performed using a densely connected UNet (DC-UNet). It is applied in two stages by considering both 2D and 3D features as input for DC-UNet, and then adding an attention mechanism to DC-UNet for better learning small tumor multi-scale features in the liver.}

\begin{figure}[!ht]
\centering
\includegraphics[width=\linewidth]{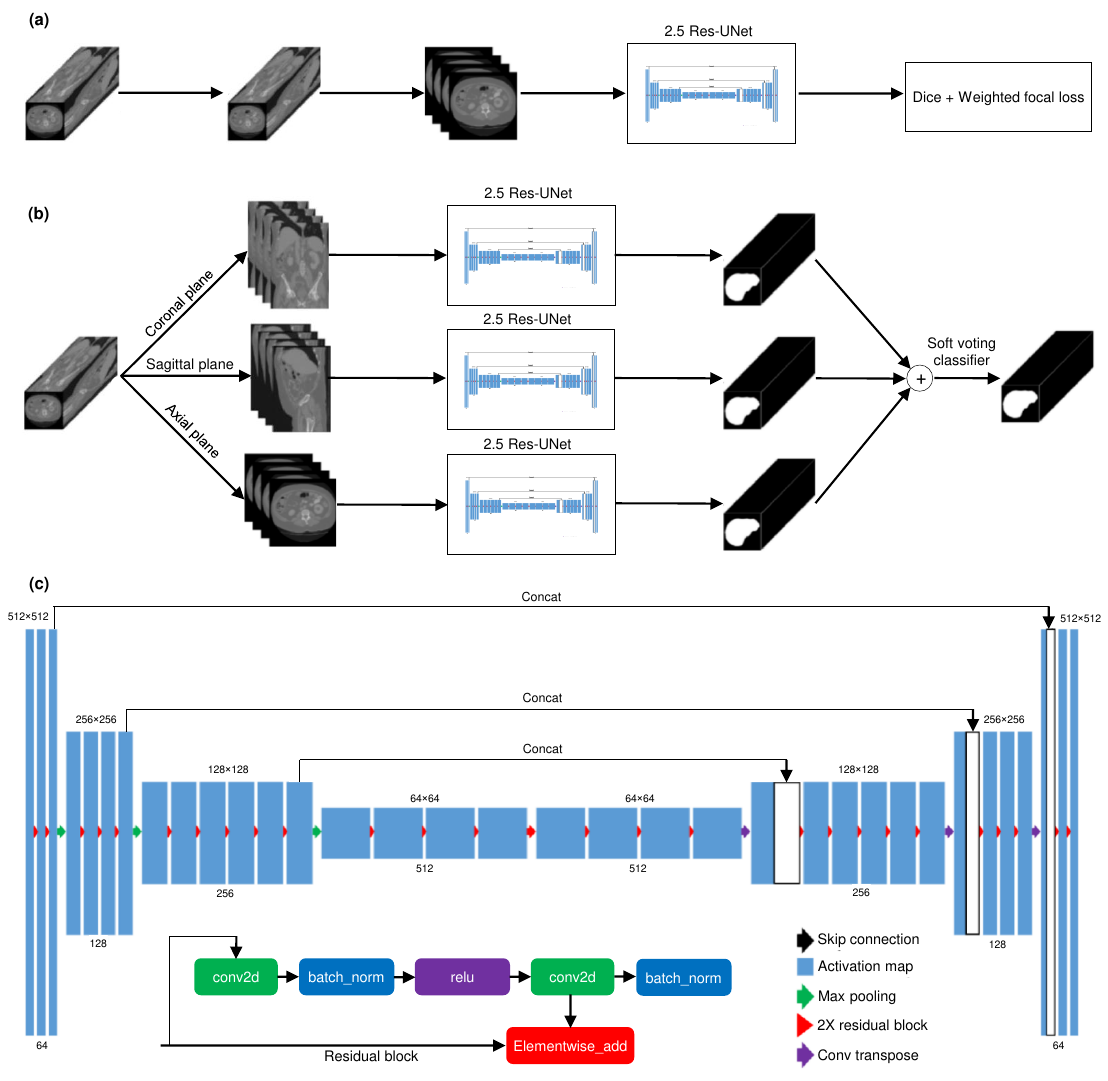}
\caption{\textcolor{black}{Flowchart of the P-UNet module proposed in \cite{han2021liver} for liver segmentation: (a) training pipeline, (b) inference pipeline, and (c) network architecture for the Res-UNet model.}}
\label{fig:P-UNet}
\end{figure}

\textcolor{black}{
In \cite{wang2021sar}, automatic liver segmentation from CT images is performed using a squeeze-and-excitation block and atrous spatial pyramid pooling based residual U-Net (SAR-U-Net). The attention strategy has been introduced to derive image features adaptively. Moreover, to extract richer multi-scale characteristics, the transition layer and the final output layer of the U-Net decoder are replaced with ASPP. Lastly, a residual block is used instead of the standard convolutional layer of U-Net before attaching a batch normalization layer to speed up the convergence.
Aiming at reducing model sizes and increasing segmentation performance, Han et al. \cite{han2021liver} fuse the output of three 2.5D Res-UNet models to develop a Perpendicular-UNet (P-UNet) for liver and hepatic tumor segmentation. Post-processing, loss functions, and data augmentation are considered to enhance the overall performance. This results in an accuracy of 96.2\% along with a DSC of 73.5\%. 
The flowchart of the developed P-UNet model is portrayed in Fig.~\ref{fig:P-UNet}, where the size has been reduced and augmented with larger receptive fields. 
}

\textcolor{black}{
A two-path CNN (TPCNN) scheme to segment tumor and liver in CT images using two encoding techniques is proposed in \cite{aghamohammadi2021tpcnn}. Typically, this scheme distinguishes the exact borders of the liver and tumors. A first encoding that faithfully extracts the necessary local shape details and increases the malleability of borders' detection with shape variation is developed. This is used even if few samples of training images exist. After that, a second encoding scheme (Z-Score normalization) is introduced to improve the distinction capability of touching organs. Lastly, a segmentation scheme using the TPCNN architecture is implemented, which relies on analyzing local and semi-global characteristics.
In \cite{araujo2022liver}, the authors proposed a liver segmentation scheme using CT images and a U-Net, which is employed in a cascaded manner. It adopts a powerful segmentation approach to segment the liver even in the presence of lesions. Typically, this framework has the advantages of (i) reducing CT examination to a region that contains the liver and initial liver segmentation; (ii) using a U-Net-based reconstruction stage for recovering liver regions affected by lesions not included in the initial segmentation; and (iii) reducing the $FPR$ and filling holes to enhance segmentation based on post-processing.
In \cite{ahmad2022lightweight}, a lightweight CNN model that reduces the computation cost of extracting the liver regions from CT scan images is proposed. The CNN model includes three convolutional and two fully connected layers. To discriminate the liver from the background, softmax has been utilized. Moreover, weight initialization has been made using random Gaussian distribution to achieve distance-preserving-embedding of the information.}

\vskip2mm
\noindent\textbf{\textit{3D FCN}} 

To further involve the volumetric information, some researchers opt for the 3D FCN, bearing in mind that the use of 3D FCN is faced with  expensive memory and computational necessities.

One of the early works of using 3D FCNs is \cite{Lu2016}, whereas a 3D FCN is utilized to segment the liver and is aided by a GC algorithm. In \cite{Hu2016,Hu2017}, the authors base their implementation on the 3D FCN employed in \cite{Lu2016}, increasing the kernels' size and modifying some of the activation functions. Then, the 3D FCN output is incorporated in an energy function optimized globally. In \cite{Dou2016,Dou2017}, a 3D FCN is equipped with a deep supervision mechanism creating a 3D deeply supervised network (3D DSN) aided by 3D CRF to refine the segmentation output. In \cite{Tang2020a}, the DeepMedic network, which relies on 3D CNN and 3D CRF \cite{kamnitsas2017efficient}, is used to segment the liver for selective internal radiation therapy (SIRT). In \cite{Gibson2017}, a 3D FCN with dilated convolutional layers is developed for multi-organ segmentation, including the liver. Following their work in \cite{Gibson2017}, another 3D FCN based on Dense V-Net is subsequently developed to segment multi-organs, including the liver \cite{Gibson2018}. The work is extended to create a framework called NiftyNet on Python \cite{Gibson2018a}, which is intended to make it easier to deploy biomedical segmentation algorithms. Another work utilizing Dense V-Net is \cite{Chung2020}, where a deeply self-supervised scheme based on adaptive contour features is utilized for the liver segmentation task alone. In \cite{Liu2018}, a 3D anisotropic hybrid network (3D AH-Net) transforms 2D weights trained on a 2D encoder into their 3D counterpart, which is then used for the liver segmentation task. In a model called Sensor3D, a 3D cascaded convolutional long short-term memory (C-LSTM) in a U-Net architecture theme is used for the segmentation task in \cite{Novikov2019}.

To further incorporate the global context information, \cite{Rickmann2019} developed project and excite (PE) modules and employed them within a 3D FCN. A similar approach is utilized in \cite{Qayyum2020}, where spatial squeeze and channel excitation (cSE) 3D modules are aiding a 3D FCN, based on ResNet architecture, in the liver segmentation task. Moreover, in \cite{Jiang2019}, a 3D FCN composed of multiple attention hybrid connection blocks, hence the name (AHCNet), has densely-connected long and short skip connections and soft self-attention modules where two cascaded AHCNets are used for liver localization and segmentation, respectively. In \cite{Takenaga2019}, to take advantage of the multi-phase volumes obtained by MRI, a multi-channel 3D FCN based on ResNet (MC-FC-ResNet), or 4D FC-ResNet, is used to segment the liver, utilizing the information each phase provides.

Needless to say, the U-Net architecture also befits in the 3D context \cite{Cicek2016}. In \cite{Bai2019}, a 3D U-Net segments the liver organ, while in \cite{Kakeya2018}, a 3D U-JAPA-Net model has a generalized 3D U-Net and a specialized one for each organ. To segment organs from different modalities, a 3D universal U-net (3D U\textsuperscript{2}-Net) is built where domain-specific convolution layers are used for each modality, and a single pipeline of convolution layers is shared across different ones \cite{Huang2019}. In \cite{Zhang2018liver}, 3D patches are inserted into a 3D U-Net-like network with context-aware units for multi-phase MRI volumes in a multi-scale fashion. In \cite{Mohagheghi2020}, 3D patches were also used; however, some modifications were applied to the 3D U-Net architecture on the pooling layers, activation functions, and channels' depth. Moreover, the work takes advantage of convolutional denoising autoencoders (CdAE) to create shape-prior knowledge and embeds it into a deep data-driven loss (DDL) to enhance the segmentation result. Finally, a 3D U-Net with a multi-scale pyramid-like liver segmentation scheme is employed in \cite{Roth2018miccai}, where it is extended in \cite{Roth2018article} to segment 20 organs in total via TL from the original segmented 8 organs in the abdomen.

It is worth mentioning that some studies use FCNs as a complementary part to the core algorithm. In the case of \cite{Zeng2019}, a mean shape fitting (MSF) algorithm, which creates an average shape of the liver, is complemented by a 3D FCN that generates a dense deformation field via the calculation of a 3D vector of displacements for each voxel to deform the created prior as necessary. Another work analyzes the inserted CT scans via 3D FCN based on DenseVoxNet, and deforms an initial sphere mesh through the use of 3D graph convolutions-based ResNet (G-ResNet), creating an elegant and smooth 3D mesh representation of the liver \cite{Yao2019}.\\

\noindent\textbf{\textit{Hybrid}} 

In \cite{Jin2018}, a 2D residual attention-aware U-Net (RA-UNet) coarsely segments the liver, which is then fed to a 3D RA-UNet counterpart to finely segment it. In \cite{Li2018}, both 2D and 3D DenseU-Net models, constituting a hybrid DenseU-Net (H-DenseUNet), are used for liver segmentation. In \cite{Zhou2020}, a redesigned U-Net model, called U-Net++, creates an ensemble mechanism from within the architecture itself, allowing the customizability of having DenseU-Nets at various levels. It is also supported by the deep supervision technique, thus, generating outputs at all levels, which then serve as ensemble models. Quickly after that, in \cite{Xu2020}, the U-Net++ model is slightly modified and used for both liver segmentation and registration between pre-operative MRI and intra-operative CT. In \cite{Li2020}, the attention mechanism and nested U-Net (ANU-Net) builds over the 2D version of the U-Net++, where modifications are applied to the loss function and the dense connections between the nested convolutional blocks.

In \cite{Zhou2019prior}, a prior-aware neural network (PaNN) single-handedly segments the liver, among other organs, trained over partially labeled datasets, similar to the training scheme deployed in \cite{Fang2020multi}. The network's 2D and 3D versions are tested and compared with other available networks.

In \cite{Zhou2019genesis}, Models Genesis is a framework that can create a basis for TL to any other organ segmentation via self-supervised training on unlabeled data instead of relying on ImageNet trained weights. The motivation is that the ImageNet dataset is different than the biomedical ones, creating an inappropriate TL process. In their work, both the 2D and 3D models are initially trained on unlabeled data and then transferred for application-specific biomedical segmentation tasks.

From Fig.~\ref{fig:liver_map}, it is obvious that the ML supervised algorithms of favor are the ones utilizing FCN as the main model, where they have become robust to tackle many problems (localization, registration, classification, or segmentation) in many fields. Moreover, within the FCN models, the majority of works have utilized 2D network models with inclusion techniques for volume information such as 2.5D inputs. The 2D models are usually preferred over 3D ones due to the aforementioned issues of expensive computations and memory shortage. \textcolor{black}{Table~\ref{tab:liver_seg} highlights the most prominent studies that have tackled open-access datasets for the liver segmentation issue, highlighting the DSC score that each study achieved, and if not available VOE is reported. The comparison has been conducted in terms of the deployed ML model, method description, dataset, best
performance, and advantage/limitation.}

\begin{small}
\color{black}
\scriptsize
\begin{longtable}{
m{0.5cm}
m{1.5cm}
m{5cm}
m{1.8cm}
m{2.5cm}
m{4cm}}
\caption{\textcolor{black}{Summary of the Liver segmentation techniques based on DL models.}}  \label{tab:liver_seg} \\
\hline
Work & ML model & Method description & Dataset & Best performance* & Advantage/limitation \\ \hline
\endfirsthead

\multicolumn{6}{c}{{Table \thetable\ (Continue)}} \\
\hline
Work & ML model & Method description & Dataset & Best performance* & Advantage/limitation \\ \hline
\endhead

\hline
\endfoot

\multicolumn{6}{l}{\rule{0pt}{4ex}*For the \enquote{Best performance} column, we mention DSC and VOE, respectively, if one is unavailable. DSC is Dice per case}
\endlastfoot
\cite{Gibson2018} & 3D FCN based on Dense V-Net & 3D DenseVNet to segment multiple organs from abdominal CT using 3D patches & BtCV \& Pancreas-CT & DSC=95\% & Multi-organ segmentation in the abdominal area, with slight enhancement on their previous work in \cite{Gibson2017} \\ \hline

\cite{Gibson2017} & 3D FCN & 3D FCN DL algorithm for liver, pancreas, stomach, and esophagus segmentation using dilated convolution units & BtCV \& Pancreas-CT & DSC=93\% & One of the early attempts to employ dilated convolutional layers for liver segmentation \\ \hline

\cite{Hu2016}, \cite{Hu2017} & 3D FCN based on \cite{Lu2016} & 3D organ delineation using 3D FCN with 10 layers, augmenting training data, and taking advantage of modern convex optimization techniques & SLIVER07 \& Private & DSC=96.0\textpm1.5\% & Absence of multi-organ datasets at the time of conducting the study \\ \hline

\cite{Tian2019} & GLC-Unet & 2xGLC-UNets to segment liver firstly and then the famous Couinaud segmentation & MSDC-T8 & DSC=98.18\textpm0.85\% (liver), DSC=92.8\textpm.08\% (Couinaud liver) & Published, liver masks (443 records) and Couinaud masks (193 records) for MSDC-T8 dataset. Also, first work to tackle Couinaud segmentation \\ \hline

\cite{Wang2019pairwise} & CoFCN & CoFCNs where two inputs are segmented in parallel and share features along segmentation & LiTS & DSC=96.43\% & Uses a portion of the dataset, not all of it \\ \hline

\cite{Zheng2019} & 2D DeepLab + DAP & Semi-supervised adversarial learning model with Deep Atlas Prior (DAP) to improve the accuracy of liver segmentation in CT images & LiTS & DSC=95.23\% & Semi-supervised model includes unannotated data in the training dataset to minimize annotation of medical images \\ \hline

\cite{tian2021fully} & 2.5D ResU-Net & Automatic annotation of liver functional regions using a 2.5D class-aware DNN & MSDC & DSC=88.2\% (liver) & Moderate computational cost with accurate segmentation. Further investigations on other datasets are needed. \\ \hline

\cite{Cheema2019} & LER-CN & Using LER-CN to segment liver from low-dose CT scan. It is achieved by two paired symmetric layers: convoluted noise removal component for coarse extraction and de-convoluted spatial preservation component for fine extraction & SLIVER07 \& Private & 92.1\textpm3.4\% (SLIVER07) & Focuses on the Low-dose CT scans (LDCT) \\ \hline

\cite{Roth2018article}, \cite{Roth2018miccai} & 3D U-Net & 3D patches segmentation using 3D multi-scale pyramid-like 3D U-Net-based model. Two-stage needed for coarse-to-fine segmentation & Pancreas-CT, VISCERAL Anatomy3 \& Private & DSC= 94.9\% (liver) & Usage of a single model (consisting of multiple networks) to segment multiple organs is a huge advantage \\ \hline

\cite{Fang2020deep}, \cite{Fang2020multi} & MIMO-FAN & A novel pyramid-like architecture for multi-organ segmentation using multi-scale fusion layers & BtCV, LiTS, KiTS \& Spleen & DSC=95.9\% (on all datasets) & The input is inserted in different scales \\ \hline

\cite{Huang2018} & K-means + modified GC & Liver segmentation based on modified GC and feature detection, which relies initially on k-mean clustering & SLIVER07 \& 3D-IRCADb & VOE=5.3\% (SLIVER07), VOE=8.6\% (3D-IRCADb) & Small tumors are under-segmented, GC-based algorithms are not suitable for elongated structures \\ \hline

\cite{Cai2019} & FKFCM & Use of FKFCM for liver and tumor segmentation & SLIVER07 \& 3D-IRCADb01 & DSC=87.02\% (SLIVER07) & High results given that unsupervised learning is used \\ \hline

\cite{Norajitra2017} & 3D SSM + RF & Liver segmentation using 3D SSM where RF utilization appears in omni-directional landmark search & SLIVER07 \& BtCV & VOE=5.90\% (SLIVER0\&), DSC=94.7\% (BtCV) & Needs flexible shape prior modeling where a case-by-case model should be used, which will increase the complexity \\ \hline

\cite{He2016} & AdaBoost + ASM & Three-level AdaBoost-guided active shape model for liver segmentation in transversal, sagittal \& coronal views & 3D-IRCADb01, SLIVER07 \& VISCERAL Anatomy3 & DSC=96.4\% (SLIVER07), DSC=93.3\% (Anatomy3) & Manual feature engineering highly dependent on the CT modality \\ \hline

\cite{Qin2018} & SLIC + 2D FCN & superpixel-based and boundary sensitive CNN for liver segmentation & LiTS & DSC=97.31\textpm0.36\% & Using superpixels as input to the FCN \\ \hline

\cite{Wang2019automatic} & 3D MPNet & Use of MPNet for liver segmentation, where each network specializes in a view and the output is the weighted-average of the 3 views based on their resolutions & LiTS & DSC=96.7\% & Multi-plane segmentation and fusion \\ \hline

\cite{Yang2019} & VAE + GAN & Usage of VAE and GANs to transfer content information to common subspace between MRI and CT & LiTS \& Private (MRI) & DSC=81\textpm3\% & VAE to disentangle information from MRI and CT modalities to find shared content space \\ \hline

\cite{Tang2020} & Faster R-CNN & Sequential segmentation of liver over two stages using R-CNN (localize) and DeepLab (segment) & SLIVER07 \& 3D-IRCADb & VOE=5.06\% (SLIVER07), VOE=8.67\% (3D-IRCADb) & Liver detection and then segmentation is interesting to reduce the work needed for the 2nd model \\ \hline

\cite{Ahmad2019} & DBN & Using DBN for liver segmentation & SLIVER07 \& 3D-IRCADb-01 & DSC=94.8\textpm0.6\% (SLIVER07), DSC=91.83\textpm1.37\% (3D-IRCADb) & The DSC dropped under 3D-IRCADb01 dataset. Also, further investigations on other datasets are needed. \\ \hline

\cite{ma2021liver} & VNet\_WGAN & Fusion of VNet and WGAN for liver Segmentation & LiTS \& CHAOS & Accuracy=94\%, DSC=92\% & Moderate performance and the loss function part is appropriate medical image analysis. \\ \hline

\cite{tan2021automatic} & 3D CNN & Automatic liver segmentation using 3D CNN and a hybrid loss function & SLIVER07 \& CHAOS & DSC=83.02\% (CHAOS) & Can be trained on small datasets; however, moderate performance is reported. \\ \hline

\cite{wang2021sar} & SAR-U-Net & Liver segmentation using an improved U-Net scheme & LiTS \& SLIVER07 & DSC=97.3\% & High computational cost compared to the state-of-the-art. \\ \hline

\cite{araujo2022liver} & U-Net & Liver segmentation from CT images using cascade DL & LiTS & DSC=95.64\% & U-Net parameters are defined empirically and liver contours of this segmentation approach have small failures. \\ \hline

\cite{ahmad2022lightweight} & Lightweight CNN & A lightweight CNN architecture to segment liver in CT images & SLIVER07, 3D-IRCADb01 \& LiTS & DSC=95\% (SLIVER07) & The accuracy dropped under 3D-IRCADb01 dataset. \\ \hline

\cite{Dou2016}, \cite{Dou2017} & 3D DSN + 3D CRF & Deeply supervised 3D CNN using auxiliary outputs + 3D CRF for contour refinement for liver, heart and great vessels segmentation & SLIVER07 \& HVSMR & VOE=5.42\textpm0.72\% (SLIVER07) & Performing well on both heart and liver segmentation, but not reporting DSC \\ \hline

\cite{Tang2020a} & DeepMedic (3D CNN + 3D CRF) & Multi-scale segmentation of the liver, with fusion of fully connected layers using DeepMedic CNN & SLIVER07, LiTS \& Private & DSC=94\% median (on Private) & Inter-observer variability was reduced when the CNN segmentation was used as a baseline \\ \hline

\cite{Chung2020} & 3D Dense V-Net & Deeply self-supervised 3D Dense V-Net to segment liver & SLIVER07, 3D-IRCADb, \cite{gibson_eli_2018_1169361}'s \& Private & DSC=96\textpm1\% & Does not investigate tumor segmentation \\ \hline

\cite{Novikov2019} & Sensor3D (C-LSTM) & Sensor3D, a bi-directional C-LSTMs following a U-Net architecture segmentation for the liver & 3D-IRCADb \& CSI2014 & DSC=95.4\% & One of few studies to use LSTM for liver segmentation \\ \hline

\cite{Rickmann2019} & 3D FCN + PE module & 3D patches inserted into 3D U-Net implementing PE modules which efficiently replaces extra convolutional layers & VISCERAL Anatomy3 \& MALC & VISCERAL: DSC=93.1\% (liver) & PE blocks can be inserted into 3D FCNS without huge effects on the computational complexity \\ \hline

\cite{Huang2019} & 3D U\textsuperscript{2}-Net & 3D patches from different CT and MRI modalities inserted into 3D Universal U-Net, for intra- and inter-modal organ segmentation & Heart, Liver, MSDC & DSC=93.54\% (liver) & Tackles both CT and MRI modalities \\ \hline

\cite{Mohagheghi2020} & 3D U-Net & Usage of 3D U-Net with hybrid loss to encompass 3D liver global knowledge for liver segmentation & SLIVER07 & DSC=97.62\% & Model's training initiates for noise removal. Would be better to test on more recent public datasets \\ \hline

\cite{Yao2019} & 3D G-ResNet & 3D Liver segmentation based on Graph-ResNet used with a backbone FCN such as 3D U-Net, V-Net, or VoxDenseNet & MSDC-T3 & DSC=96.47\% & The study also generates 3D measure of the organ that is segmented \\ \hline

\cite{Zhou2020} & U-Net++ & Revolutionized U-Net architecture with deep supervision for different organs' segmentation where the encoder is redesigned to support inter-level concatenations & LiTS \& Others & DSC=82.6\textpm1.11\% & A part of the model can be considered a standalone U-Net, and the output of the U-Net++ is the ensemble of all those existing U-Nets within \\ \hline

\cite{Li2020} & 2D ANU-Net (based on U-Net++) & Modifications to the U-Net++ architecture and loss function for multi-organ segmentation & LiTS \& CHAOS & DSC=98.15\% (LiTS), DSC=93.55\% (CHAOS) & Work introduces a light ANU-Net version, with a reduction in performance \\ \hline

\cite{Zhou2019prior} & PaNN & PaNN network in two versions (2D \& 3D) for organ segmentation using partially annotated datasets for different organs' segmentation & BtCV, MSDC-T9, Pancreas-CT \& MSDC-T3 & DSC=97.4\% & Multi-organ segmentation, where the model also performs well for pancreas segmentation \\ \hline

\cite{Zhou2019genesis} & Genesis & Using self-supervised training for a model where this model is used as a basis to train for a specific biomedical task (ImageNet biomedical counterpart) & LiTS & DSC=91.13\textpm1.51\% & Interesting self-supervised training technique to teach the model human anatomy \\ \hline

\end{longtable}
\end{small}

\subsubsection{Tumors/Lesions}
\begin{figure}[!ht]
    \centering
    \includegraphics[width=\linewidth, trim={0in 3in 0in 0in},clip]{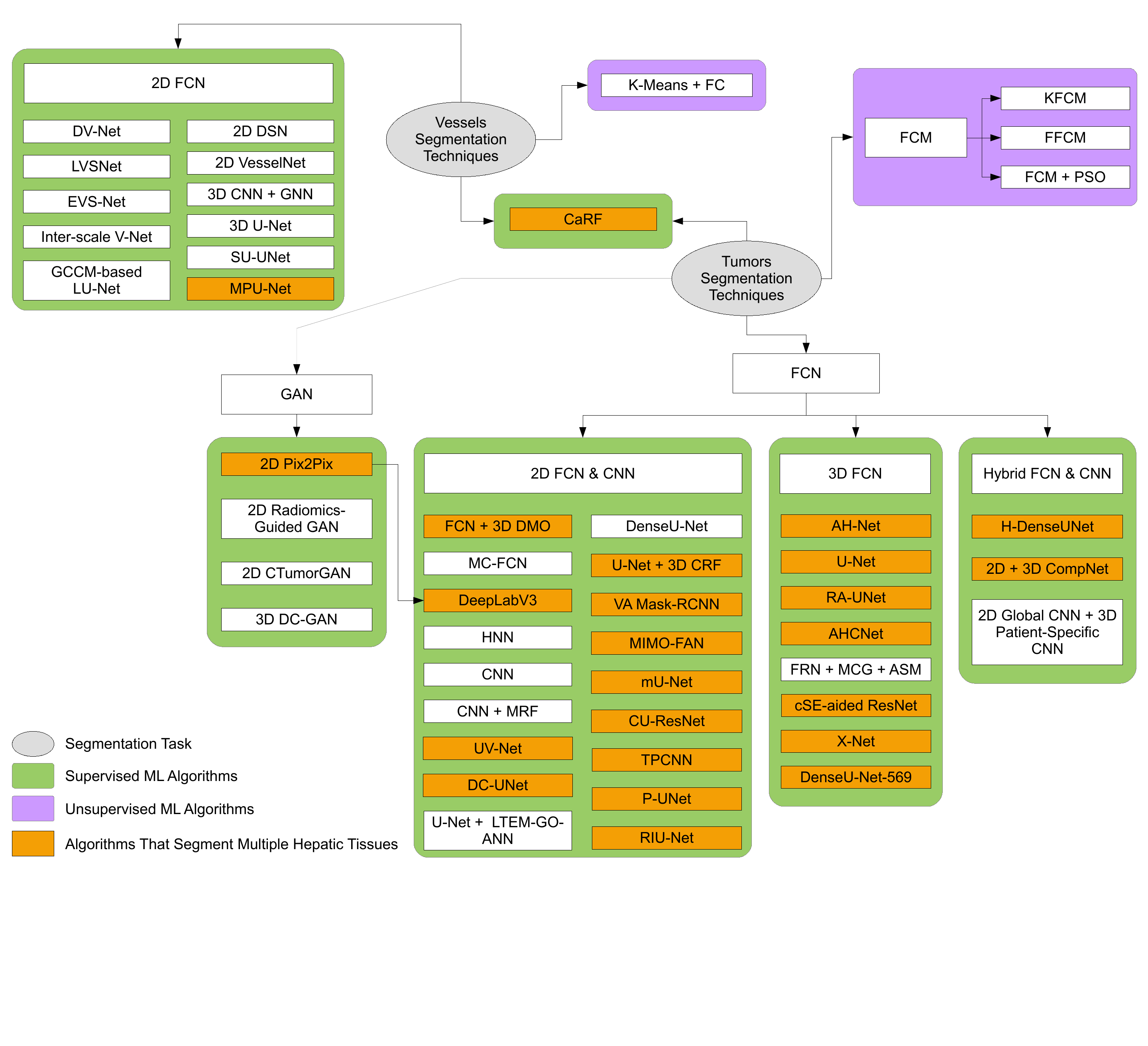}
    \caption{Tumors and vessels segmentation techniques from the reviewed literature}
    \label{fig:tumors_and_vessels_map}
\end{figure}

\textcolor{black}{The second tissue of importance within the liver is the tumors, where many datasets have focused on the idea of detecting and segmenting the existing tumors within the liver. Fig.~\ref{fig:tumors_and_vessels_map} have been created, detailing different ML algorithms, supervised and unsupervised, tackling the issue of tumors tissue segmentation. it is worth noting that there are notable intersections that are highlighted (using a light brown color) among the three classified tissues within the liver between Fig.~\ref{fig:liver_map} and Fig.~\ref{fig:tumors_and_vessels_map}.}

To segment tumors, \cite{Das2016} initially uses a Kernelized FCM (KFCM), then utilizes spatial-FCM in \cite{Das2019} for the tumor segmentation task, followed by a 4.5C decision tree (DT) algorithm to classify segmented tumors. In \cite{Anter2018}, the combination of PSO and FFCM is used for tumor segmentation, while FFCM is utilized for the tumor segmentation task in \cite{Sayed2016} and \cite{Anter2019} along with NS and adaptive watershed algorithm.

2D CNN with fully connected layers is used in \cite{Li2015} to segment tumors on patches, testing different patch sizes for optimal performance. By patching the slices, it allows the model to focus on the tumor itself instead of processing many unrelated pixels at the same time as they are normally sparse. Also, a comparison is drawn against other ML algorithms such as AdaBoost, RF, and SVM, proving the superiority of CNN-based techniques. In \cite{Sun2017}, a single 2D FCN on single-phase CT and 2D MC-FCN utilizing three phases of CE-CT scans are utilized for segmenting tumors within the liver, where the networks are initially trained on the liver segmentation task to allow for faster convergence when trained on the tumor segmentation one. In \cite{Chlebus2018}, a DenseU-Net is utilized for the tumor segmentation task, where post-processing for object identification is based on RF to reduce FPs. In \cite{Vivanti2017}, a Markov random field (MRF) registration technique is used to delineate the tumors in a follow-up CT scan from a baseline one. Then, a 2D CNN is used to segment new tumors, an RF is also used for tumor classification. Following their work, in \cite{Vivanti2018}, a global 2D CNN and patient-specific 3D CNN are used to segment the tumors on the follow-up CT scans, where if the global CNN achieved low results, the patient-specific CNN is opted for.

An interesting approach is investigated on imprecise labeling of tumors, named ``response evaluation criteria in solid tumors'' (RECIST). Due to the abundance nature of this kind of data, in \cite{Cai2018}, the authors use such CT slices for the tumor segmentation via a holistic nested network (HNN), which is originally built for edge detection throughout multiple levels within the network \cite{Xie_2015_ICCV}. Thus, utilizing weakly labeled data instead of relying on pixel-wise labeling. In a similar approach in \cite{Cano-Espinosa2020}, a 2D FCN is used to regress biomarker information (area or volume) on CT slices to regress and localize tumors instead of using manually labeled-pixels datasets, which are harder to obtain. 

Some of the aforementioned liver segmentation work also use the same model for tumor segmentation. For example, the 2D FCN aided by the 3D DMO and NMF in \cite{Zheng2017}, the 2D FCN in \cite{Tian2018}, the 2D FCN VA Mask-RCNN in \cite{Wang2019volumetric}, the TPCNN in \cite{aghamohammadi2021tpcnn}, the DeepLabV3 followed by Pix2Pix GAN in \cite{Xia2019}, the ensemble of the three U-Net-like 2D FCN in \cite{Vorontsov2018,Maaref2020}, the mU-Net in \cite{Seo2020}, the CU-ResNet \cite{Xi2020}, the 3D AH-Net in \cite{Liu2018}, the 3D U-Net with context-aware modules in \cite{Zhang2018liver}, and the H-DenseUNet in \cite{Li2018}, all segment the tumors along with the liver simultaneously. In \cite{Dey2020}, the first 2D CompNet, aforementioned in the liver segmentation techniques, helps in segmenting large tumors, followed by a 3D CompNet to segment the smaller ones. In \cite{Treilhard2017}, the same CaRF used for liver parenchyma segmentation is also used for viable tumor tissue and necrosis tissues segmentation.  In \cite{Li2021}, the same U-Net along with the Bi-ConvLSTM are used to finely capture the tumors borders. 

In contrast, some of the works opt to use an extra network for the tumor segmentation task in a cascaded sequence. In \cite{Christ2016,Christ2017}, another 2D FCN is used to segment the tumors, within the segmented liver from the first 2D FCN, in a CFCN fashion, where the 3D CRF refines the output of the CFCN model. Similarly, in \cite{Han2017}, another DenseU-Net is used to segment tumors from the segmented liver from the first DenseU-Net. In \cite{Yuan2017}, a third 2D FCN, following the first two that segmented the liver, is used for the tumor segmentation task. Moreover, in \cite{Jin2018}, a third 3D RA-UNet segments tumors from the liver mask outputted from the first two (2D and 3D) RA-UNets networks. Similarly, a third AHCNet is used for tumor segmentation on the segmented liver from the first two AHCNets \cite{Jiang2019}. On the other hand, in \cite{Ouhmich2019}, after segmenting the liver with a single U-Net, two other cascaded U-Nets are used for tumor, and its viability segmentation, respectively. In \cite{Nanda2019}, the authors opt for a different network than the SegNet employed for segmenting the liver. Laws texture energy measure (LTEM) features are extracted for tumor detection using a normal ANN optimized by a genetic optimizer (GO) algorithm (LTEM-GO-ANN), followed by a 2D U-Net performing the tumors segmentation on detected tumor regions. Similarly, in \cite{Chen2019a}, after segmenting the liver using the MPNet, a 3D densely-connected GAN (DC-GAN) is used for the tumor segmentation within the segmented liver. In \cite{Bai2019}, after segmenting the liver using 3D U-Net, a multi-scale candidate generation (MCG) algorithm generates candidate tumor areas based on superpixels, which are inserted into a 3D fractal residual network (FRN), and the output is refined by an ASM algorithm. In \cite{Jansen2019}, after segmenting the liver using a 2D MC-FCN, another network with dual-pathways is used to segment the tumors using nine phase slices. In \cite{Zhang2020a}, after coarsely segmenting the liver using a 2D U-Net, a 3D FCN is used to segment the tumors within, followed by an LSM algorithm to refine the tumor segmentation. In \cite{Araujo2021}, a RetinaNet comes before the U-Net to create an initial segmentation for the tumor that is then utilized by the U-Net to complete the task of tumors segmentation, followed by some post-processing techniques to enhance the final segmentation results.

GANs are also used in the tumor segmentation task. In \cite{Xiao2019}, a radiomics-guided GAN utilizes a dilated DenseU-Net as the generator (segmenter), and a VGG network as the discriminator in GAN where the discriminator extracts radiomics features to aid the segmenter in tumor segmentation. However, in \cite{Pang2020}, a 2D CTumorGAN is used for the tumor segmentation task in multiple organs, including the liver, and it incorporates a novel generator scheme that integrates a noise vector with the encoder part to generate segmentation masks.

\textcolor{black}{Similar to Table~\ref{tab:liver_seg},  Table~\ref{tab:tumor_seg} highlights the studies that focus on the task of tumor segmentation, along with liver segmentation if the same work segments both. They are compared in terms of the deployed ML model, method description, dataset, best performance, advantage/limitation, and segmentation category.}

\begin{small}
\color{black}
\scriptsize
\begin{longtable}{
m{0.5cm}
m{1.5cm}
m{3.5cm}
m{1cm}
m{3.5cm}
m{4cm}
m{1cm}}
\caption{\textcolor{black}{Summary of the tumors (and liver) segmentation frameworks based on ML and DL models.}} \label{tab:tumor_seg}\\
\hline
Work & ML model & Method description & Dataset & Best performance* & Advantage/limitation & Segmentation category \\ \hline
\endfirsthead

\multicolumn{7}{c}{{Table \thetable\ (Continue)}} \\
\hline
Work & ML model & Method description & Dataset & Best performance* & Advantage/limitation & Segmentation category \\ \hline
\endhead

\hline
\endfoot
\multicolumn{7}{l}{\rule{0pt}{4ex}*For the \enquote{Best performance} column, we mention DSC, JI, and VOE, respectively, if one is unavailable. DSC is Dice per case}
\endlastfoot

\cite{Christ2017}, \cite{Christ2016} & U-Net-like 2D FCN + 3D CRF & Cascaded FCNs (U-Nets) for 1) liver segmentation \& 2) lesion segmentation aided by 3D CRF on both CT and MRI & 3D-IRCADb \& 2 Private Datasets & 3D-IRCADb: DSC=94\% (liver) \& DSC=56\% (tumors), Private CT: DSC=88-91\% (liver), DSC=61\% (tumors validation set), Private MRI: DSC=87\% (liver), 69.7\% (tumors) & Large dataset, possibly the studies that revolutionized liver/tumors segmentation challenge as they create the LiTS dataset & Liver \& Tumors \\ \hline

\cite{Jiang2019} & 3D AHCNet & 3D AHCNet combines soft and hard attention mechanisms with long and short skip connections for liver and tumor segmentation & LiTS \& 3D-IRCADb \& Private & 3D-IRCADb: DSC=95.3\% (liver), DSC=66.8\% (tumor), LiTS: DSC=59.1\% (tumor) & tumor segmentation can be enhanced perhaps by post-processing techniques & Liver \& Tumors \\ \hline

\cite{di2022automatic} & 3D U-Net & Using hierarchical
iterative superpixels and local statistical features for automatic liver tumor segmentation & LiTS \& 3D-IRCADb & DSC=71\% (3D-IRCADb) & Less computational cost compared existing methods. & Liver \& Tumors \\ \hline

\cite{alalwan2021efficient} & 3D DenseU-Net-569 & A fully 3D semantic segmentation network with lower training parameters and deeper network. & LiTS & DSC=96.7\% (liver),\newline DSC=80.7\% (tumor) & Validated only on one dataset and the performance of tumor segmentation needs further improvement. & Liver \& Tumors \\ \hline

\cite{Han2017} & DenseU-Net & Use of 32-layered deep CNN with short (ResNet Connections) and long connections (U-Net connections) & LiTS & DSC=67\% & Using better post-processing techniques or ensemble of FCNs & Tumors \\ \hline

\cite{Li2018} & 2D \& 3D DenseU-Net models & 2D DenseU-Net for extracting intra-slice features and a 3D counterpart for hierarchically aggregating volumetric context for liver \& tumor segmentation & LiTS \& 3D-IRCADb & DSC=96.1\% (liver), \newline DSC=72.2\% (tumor) & Small tumors are under-segmented & Liver \& Tumors \& Tumor Burden \\ \hline

\cite{Chlebus2018} & DenseU-Net + RF & Usage of multiple 2D U-Net for liver and tumor segmentation. For the tumor segmentation part, RF is used for FP filtering & LiTS & DSC=96\% (liver),\newline DSC=68\% (tumor) & Smaller tumors pose an issue for the FCN & Liver \& Tumors \\ \hline

\cite{Vorontsov2018} & U-Net-like 2D FCN & Sequential U-Net-like models for liver and tumor segmentation with an ensemble of models for the liver segmentation & LiTS & DSC=95.1\% (liver),\newline DSC=77.3\% (tumor) & Tumor segmentation is high, liver not the best, maybe increase 2.5D input size could help & Liver \& Tumors \\ \hline

\cite{zhang2021liver} & UV-Net-Multi-scale & Liver tumor segmentation using 2.5D UV-Net with multi-scale convolution & LiTS & DSC=88.92\% (liver) & Moderate training time and further investigations on other datasets are needed. & Liver \& Tumors \\ \hline

\cite{lv20222} & lightweight RIU-Net & Automatic liver and tumor segmentation from CT images by extracting inter-slice spatial information in the form of 2.5D. & LiTS \& 3D-IRCADb & DSC=97.72\% (LiTS) & Improves the performance in the presence of small tumors or tumors around the liver boundary; however, insufficient feature integration still exists. & Liver \& Tumors \\ \hline

\cite{han2021boundary} & Cascaded 2.5D FCN & Segmentation of liver and tumor by exploring spatial information in 3D images. & LiTS \& 3D-IRCADb & DSC=96.1\% (liver, 3D-IRCADb), DSC=74.5\% (tumor, LiTS) & The performance needs further improvement especially for the case of tumor segmentation. & Liver \& Tumors \\ \hline

\cite{zheng2022automatic} & 3D CNN+C-LSTM & Automatic liver tumor segmentation using 4D information & Private & DSC=82.5\% & Small tumors could be missed as the size if liver tumors is variable. Motion artifacts could generate co-registration errors between multi-phase DCE images. & Tumors \\ \hline

\cite{Jin2018} & 2D RA-Unet & 3D patches liver and tumors segmentation using cascaded RA-UNet powered by attention-aware capabilities & LiTS \& 3D-IRCADb & LiTS: DSC=96.1\% (liver), DSC=59.5\% (tumor), \newline 3D-IRCADb: DSC=97.7\% (liver), DSC=83\% (tumor) & Heavyweight 3D FCN model is used, which is more expensive to train & Liver \& Tumors \\ \hline

\cite{Dey2020} & 2D \& 3D CompNet & A hybrid of 2D CompNet (liver) \& 3D CompNet (tumors) cascaded to segment liver and tumors (large and small) & LiTS & DSC=68.1\% (tumor) & hindered by LiTS imperfect segmentation of tumors & Liver \& Tumors \\ \hline

\cite{Liu2018} & 3D AH-Net & 3D anisotropic hybrid network (AH-Net) utilizing inter-slice information to gain intra-slice features (2D to 3D) & Breast tumors (Private) \& LiTS & DSC=96.3\% (liver), \newline DSC=63.4\% (tumor) & exceeds state-of-the-art at the time, also the DPC is the one to be compared with the other DSC & Liver \& Tumors \\ \hline

\cite{Tian2018} & 2D FCN & FCN to segment CT slices, and a separate LSTM language model to generate captions creating a diagnostic report generator & LiTS & DSC=94.2\% (liver), \newline DSC=54.9\% (tumor) & Opens aspects of report generation and image captions for medical reasons & Liver \& Tumors \\ \hline

\cite{Yuan2017} & 2D FCN & 3x2D FCN cascaded for liver and tumor segmentation on LiTS dataset & LiTS & DSC=96.3\% (liver), \newline DSC=65.7\% (tumor) & A participant of the original LiTS competition, hardware limitations prevented higher yield in results & Liver \& Tumors \\ \hline

\cite{aghamohammadi2021tpcnn} & TPCNN & Tumor and liver segmentation encoding techniques in CT images & Dataset in \cite{ranjbarzadeh2020automated} & & Performance needs further improvement and validation on real-world scenarios is missing. & Liver \& Tumors \\ \hline

\cite{Nanda2019} & LTEM-GO-ANN \& 2D U-Net & Cascaded SegNet, ANN and U-Net, where former segments liver, middle detects tumors, and latter segments tumors & LiTS & DSC=69.76\% & Localizes tumors before segmenting it & Liver \& Tumors \\ \hline

\cite{Zhang2020a} & 2D U-Net \& 3D FCN + LSM \& FCM & Deep learning methods for liver and tumor segmentation aided by level-set method & LiTS \& Private & DSC=96.3\% (liver),\newline DSC=71.8\% (tumor) & Utilizing FCM to help FCN in tumor localization is interesting & Liver \& Tumors \\ \hline

\cite{Seo2020} & mU-Net & Modifications to the U-Net skip connection architecture & LiTS \& 3D-IRCADb & DSC=98.51\textpm1.02\% (liver), DSC=89.72\textpm5.07\% (tumor) on LiTS & No preprocessing is required & Liver \& Tumors \\ \hline

\cite{Chen2019a} & MPNet + ADCN & Use of MPNet to segment the liver followed by an ADCN to segment the tumor & LiTS & DSC=96.7\% (liver),\newline DSC=68.4\% (tumor) & Multi-plane segmentation and fusion & Liver \& Tumors \\ \hline

\cite{Xi2020} & CU-ResNet & Comparison between loss functions on CU-ResNet on both liver and tumors segmentation challenges & LiTS & DSC=94.9\% (liver),\newline DSC=75.2\% (tumor) & Fine-tune the weighting parameters in the loss function could enhance the results & Liver \& Tumors \\ \hline

\cite{Wang2019volumetric} & VA Mask-RCNN (2D FCN) & Using volumetric attention modules with CNN to enhance the extracted inter-slice information in 2.5D-based segmentation & LiTS \& DeepLesion & LiTS: DSC=74.1\% (tumor) & VA modules can be integrated with any type of CNN & Liver \& Tumors \\ \hline

\cite{Xia2019} & DeepLabV3 + Pix2Pix GAN & Usage of Pix2Pix GAN after DeepLabV3 network to improve liver and tumor segmentation \& detection & LiTS \& DeepLesion & LiTS: DSC=97.0\% (liver) & Creating a weighted complex loss function consisting of multi-class cross entropy, generator and discriminator loss functions & Liver \& Tumors \\ \hline

\cite{meng2021two} & DC-UNet & Two-stage liver and tumor segmentation using 2D and 3D features and attention mechanism & LiTS & DSC=96.7\% (liver) \newline DSC=72.5\% (tumor) & The performance of tumor segmentation needs further improvement. & Liver \& Tumors \\ \hline

\cite{han2021liver} & P-UNet & Liver and hepatic tumor segmentation by fusing the outputs of three perpendicular 2.5D Res-UNets. & LiTS & Acc=96.2\%, DSC of 73.5\% & Trains with less data and GPU memory; however, it is only trained on one dataset. & Liver \& Tumors \\ \hline

\cite{Qayyum2020} & 3D ResNet-based FCN + cSE modules & Segmentation of liver and kidneys and their associated tumors via 3D ResNet and cSE blocks & LiTS \& KiTS & DSC=97.1\% (liver),\newline DSC=82.5\% (tumor) & Tumor segmentation results are superb & Liver \& Tumors \\ \hline

\cite{Bai2019} & 3D U-Net + MCG + 3D FRN + ASM & 3D U-Net for liver segmentation followed by 3D FRN for tumors segmentations followed by ACM for refining the tumor segmentation & LiTS & DSC=67.4\% (3D-IRCADb & The study notes that 3D-IRCADb is part of the LiTS, and thus, that portion should not be used for training & Tumors \\ \hline

\cite{Sun2017} & 2D FCN \& 2D MC-FCN & Usage of FCNs on 3D-IRCADb and MC-FCNs on JDRD for tumors segmentation & 3D-IRCADb \& JDRD & VOE=15.6\textpm4.3\% (3D-IRCADb), VOE=8.1\textpm4.5\% (JDRD) & Segments liver from multi-phased CT images, but no mention of MRI & Tumors \\ \hline

\cite{Cai2018} & HNN & Weakly supervised segmentation approach utilizes RECIST-based lesion diameter measurements into full 3D lesion volume segmentation and measurements in different organs & lymph node (LN) dataset \& DeepLesion & DSC=92\% (RESICT tumor), DSC=76\% (tumor volumes) & Utilizes RECIST data to segment different tumors around the body including liver's & Tumors \\ \hline

\cite{Pang2020} & 2D CTumorGAN & A universal tumor segmentation using GAN network on CT scans for tumor segmentation around the body from different datasets in kidneys, lungs and liver & LiTS \& Others & DSC=80.19\% & Work concludes that MSE is the best loss function for tumor segmentation & Tumors \\ \hline
\end{longtable}
\end{small}

\subsubsection{Vessels}
Application of unsupervised-based ML algorithms towards segmentation of vessels is rare and only available in \cite{Zhang2018vessels}, where Jerman’s vesselness filter based on K-means clustering is followed by an improved fuzzy connectedness (FC) algorithm to segment the vessels. On the other hand, the only supervised-based ML study that segments all the liver's tissues, i.e., liver parenchyma, tumors, and vessels, is in \cite{Treilhard2017}, using the same aforementioned CaRF for blood vessels segmentation.

The majority of existing works employ supervised-based ML algorithms. In \cite{Zeng2016}, an anisotropic filter is used to suppress noise and simultaneously maintain boundary details. Followed by the use of the four filters: 1) Sato; 2) Frangi 3) offset medialness; and 4) strain energy to extract vessel features, which are then normalized. Finally, an extreme learning machine (ELM) is applied to recognize liver vessels from the background. In \cite{Mishra2019}, a more ML-dependent approach is followed. A 2D DSN based on VGG-16 is used on liver vessels segmentation from US imagery. DSN has three layers: a) object boundary definition prediction by fine resolution layers aided by auxiliary losses; b) coarse resolution layers to discriminate object regions within the boundary; and c) a trainable fusion layer. In \cite{Kitrungrotsakul2019}, a 2D VesselNet describes an architecture that utilizes three DenseNets aimed at segmenting orthogonal patches, pre-processed by a Frangi filter, from the three planar views (transversal, sagittal, coronal). Thus, vesselness probability maps are inserted into the orthogonal DenseNets, which are then fused to generate the final segmentation mask. To incorporate the 3D context even further, some researchers developed 3D networks. In \cite{Huang2018vessels}, a 3D U-Net is employed, which is vital in the case of tubular structures traversing narrowly through the slices. The work emphasizes the data imbalance issue and attempts to solve it by carefully designing data augmentation schemes and loss functions. In \cite{Zhang2020graph}, a 3D CNN for vessel enhancement is used to highlight the vessel centerlines. A 3D tree tracing algorithm initializes the vessel graph tracing with high sensitivity and low specificity. Then, a graph neural network (GNN) equipped with graph attention layers (GAT) is utilized to prune the false-positive branches.

\textcolor{black}{
In \cite{zeng2018automatic}, two strategies, namely (i) the 3D region growing; and (ii) the hybrid active contour model, are combined to segment liver vessels by utilizing the shape and intensity constraints of 3D vessels. The first one has been facilitated by the bi-Gaussian filter for thin vessel segmentation, while the second has been combined with K-means clustering for thick vessel segmentation. 
Besides, in \cite{su2021dv}, a dense V-Net (DV-Net) model is proposed to segment liver vessels, which relies on (i) integrating a dense block structure into V-net; and (ii) using data augmentation from abdominal CT volumes with scarce training data. Additionally, a dual-branch dense connection down-sampling strategy (DCDS) and a Dice and BCE (D-BCE) loss function have been introduced for capturing vascular features and maximizing image resources' use, respectively.
Moving forward, in \cite{hao2022hpm}, a hierarchical progressive multiscale network, namely HPM-Net, is proposed to segment liver vessels in CT images. Accordingly, the multiple-scale semantic features of liver vessels are learned using a hierarchical progressive multiscale learning model, aggregating internal and external progressive learning methods. Moreover, a dual-branch progressive 3D Unet, using a dual-branch progressive (DBP) down-sampling approach, is introduced to capture vessel features better.
}

\begin{figure}[!t]
\centering
\includegraphics[width=\linewidth]{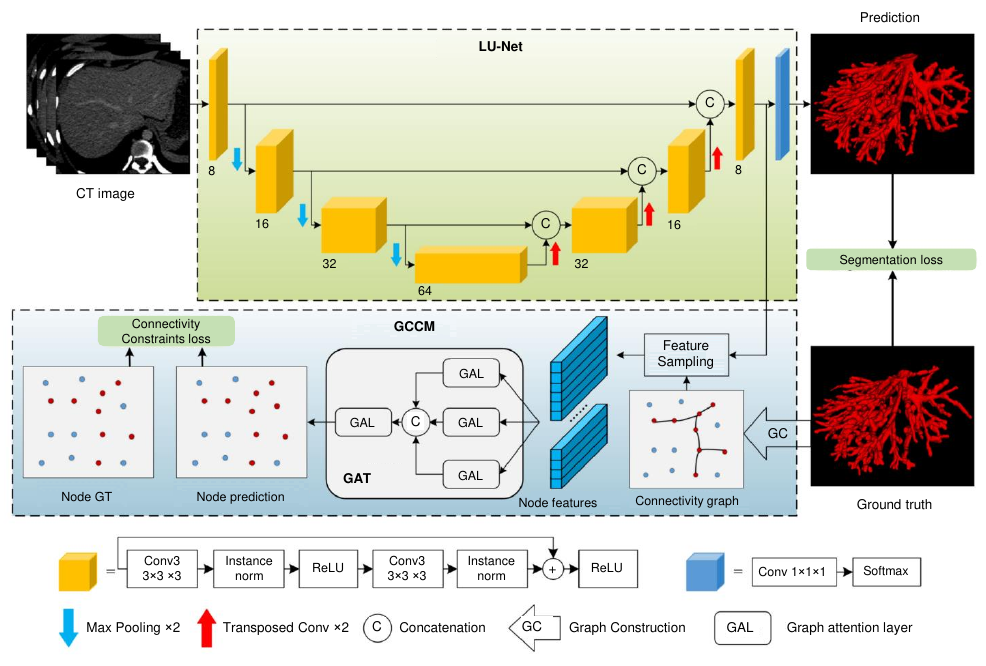}
\caption{\textcolor{black}{Flowchart of the liver vessel segmentation method proposed in \cite{li20213d}, which is conducted in two stages by (i) using a 3D lightweight LU-Net for segmentation, and (ii) applying a GAT-based GCCM to supervise LU-Net training process with \enquote{connectivity prior} information of the hepatic-vessels.}}
\label{fig:vessel_seg}
\end{figure}

\textcolor{black}{
Many liver vessel segmentation methods have relied on using UNet architecture; however, interference can be caused as not all features of the encoder are useful. To avoid this issue, Yan et al. \cite{yan2020attention} propose LVSNet, a lever vessel segmentation network that deploys particular designs to derive the accurate structure of the liver vessels. 
Typically, an attention-guided concatenation (AGC) component is designed for adaptively selecting relevant context characteristics from low-level characteristics guided by high-level characteristics. 
Put simply, the AGC component concentrates on deriving rich complemented data to get more details.
In \cite{feng2021edge}, the EVS-Net is introduced to segment the pathological liver vessels with limited labels. This module includes a pathological vessel segmentation model with two discriminators. The segmentation model has been initialized with supervision training under limited labeled patches. Simultaneously, to enhance the edge consistency of massive unlabeled patches, an edge-aware self-supervision architecture has been developed. 
In \cite{yang2021liver}, an inter-scale V-Net model is developed to segment liver vessels by (i) introducing a dilated convolution into the traditional V-Net model to help it in saving detailed spatial information and enlarging the receptive field without reducing down-sampling; and (ii) using a 3D deep supervision mechanism into the V-Net to speed up its convergence enables it to better learn semantic features. 
Additionally, to efficiently integrate multi-scale features and avoid the loss of high-level semantic information in the decoder, inter-scale dense connections are integrated into the decoder of V-Net.
In \cite{montana2021vessel}, Montana et al. utilize 2D U-Net with batch-norm blocks at each layer for liver vessel segmentation in 2D registration of laparoscopic ultrasound (LUS) images. 
}

\textcolor{black}{
In \cite{li20213d}, liver vessel segmentation is performed by introducing a new strategy, namely plug-in mode, which helps in integrating a graphical connectivity constraint module (GCCM) into a 3D lightweight U-Net (LU-Net). In this regard, the 3D lightweight LU-Net has been used for segmentation. In contrast, the GCCM, which is based on a graph attention network (GAT), has been utilized for supervising the training process of LU-Net with connectivity prior.
Because of low contrasts in the CT images and complex structures of hepatic vessels, it is challenging even for experts to perform accurate manual annotation. Consequently, most of the labels of existing publicly available datasets are noisy. Fig.~\ref{fig:vessel_seg} portrays the flowchart of the liver vessel segmentation method proposed in \cite{li20213d}.
To overcome this issue, Liu et al. \cite{liu2022unet} introduce a self-updating U-Net (SU-UNet), based on soft-constraint, for an accurate vessel segmentation from noisy annotations. Two U-Net architectures are first used to output different segmentation predictions before designing a SU module to optimize noisy vessel labels based on segmentation predictions. This helps better guide the model training using optimized labels.
Finally, Table~\ref{tab:vessel_seg} summarizes some existing liver vessel segmentation frameworks discussed in the paper. A comparison has been conducted in terms of the deployed ML model, method description, dataset, best performance, and advantage/limitation.
}

\begin{table}[!ht]
\caption{\textcolor{black}{Summary of the liver vessel segmentation frameworks based on ML and DL models.}}
\label{tab:vessel_seg}
\centering
\scriptsize
\color{black}{
\begin{tabular}{
m{0.5cm}
m{1.3cm}
m{5.2cm}
m{2.1cm}
m{1.9cm}
m{4.5cm}
m{2cm}
}
\hline
{\scriptsize Work} & {\scriptsize ML model} & {\scriptsize %
Method description } & {\scriptsize Dataset} & {\scriptsize Best performance} & 
{\scriptsize Advantage/limitation} \\ \hline

\cite{Ivashchenko2020} & 4D K-means & An automatic algorithm for liver based on 4D K-means and vessels segmentation using Otsu and Frangi filters & Private (MRI) & DSC=94.9\textpm1.2\% (liver) & Segmentation of the hepatic vein could be improved, and more common metrics could have been used \\ \hline

\cite{Treilhard2017} & CaRF & Multi-class tissue segmentation (Parenchyma, viable tumor, necrosis, and vasculature) from multi-parameter MRI & Private (MRI) & DSC=67.8\% (viable tumor), DSC=54.4\% (necrosis), DSC=55.7\% (vessels) & Private dataset, but one of the only studies to tackle all the tissues segmentation \\ \hline

\cite{Perslev2019} & MPU-Net & MPU-Net to segment multiple organs from different views with the same simple architecture & MSDC & DSC=76\textpm18\% (liver \& tumor MSDC-T3), DSC= 49\% (tumor \& vessels, MSDC-T8) & Tackles ALL challenges in MSDC \\ \hline

\cite{Zhang2018vessels} & K-means + FC & Improved FC method for automatic 3D liver vessel segmentation in CT images. & 3D-IRCADb \& SLIVER07 & DSC=67.3\textpm5.7\% (3D-IRCADb), DSC=71.4\textpm7.6\% (SLIVER07) & Small dataset size, and the isotropical and anisotropical resampling are time-consuming \\ \hline

\cite{Zeng2016} & ELM & Vessels segmentation using ELM on pre-processed data (anisotropic filter is used to suppress noise + Sato, Frangi, offset medialness, and strain energy filters to extract vessel features) & Private & Acc=98.1\%, \newline Recall=74.2\% & One of the earliest work to tackle liver vessels segmentation \\ \hline

\cite{Mishra2019} & 2D DSN based on VGG-16 & Use of deeply supervised CNN to segment vessels in the liver within US images & Private (US) & DSC=73\% & Focuses on the US images, which is rare in literature \\ \hline

\cite{Kitrungrotsakul2019} & VesselNet & A DenseNet-based 2D FCN is used to segment vessels from Fringi filtered vessels images using orthogonal 2D patches from the 3 different views & 3D-IRCADb \& ViscuSynth & DSC=90.3\% (3D-IRCARDb) & Excellent results, but 2-4 mins for computations (with GPU) \\ \hline

\cite{Huang2018vessels} & 3D U-Net & 3D vessels segmentation using 3D U-Net with an adjusted loss function & 3D-IRCADb \& SLIVER07 \& Private & DSC=75.3\% (3D-IRCADb) & High vessels segmentation DSC \\ \hline

\cite{Zhang2020graph} & 3D CNN + GNN & 3D CNN vessels' highlighting network on 64x64x64 patches followed by GNN to properly segment hepatic vessels & Private & F1 Score/DSC =~87.62\textpm5.49\% & Uses GNN as a post-processing scheme to enhance F1 Score \\ \hline

\cite{zeng2018automatic} & K-means clustering & Automatic liver vessel segmentation using 3D region growing and hybrid active contour model & Private & DSC=73\% & Some thick vessels with lower intensities fail to be segmented and the surface of segmented liver vessels is not convincingly smooth. \\ \hline

\cite{su2021dv} & DV-Net & Dense connection model with D-BCE loss function for automatic liver vessel segmentation. & 3D-IRCADb \& MSDC & DSC=75.46\% & High computational cost and the performance needs further improvement. \\ \hline

\cite{hao2022hpm} & HPM-Net & Liver vessel segmentation based on 3D CNN &  3D-IRCADb & DSC=75.18\% & The network has a complex structure withe a huge naumber of parameters. \\ \hline

\cite{li20213d} & GCCM-based LU-Net & Liver vessel segmentation using 3D graph-connectivity constrained network &  3D-IRCADb \& MSDC & DSC=65.41\% (3D-IRCADb) & The GPU memory consumption increases by decreasing sampling interval. By contrast, increasing the sampling interval results in lower segmentation performance. \\ \hline

\cite{yan2020attention} & LVSNet & Attention-guided DL with multi-scale feature fusion for liver vessel segmentation & 3D-IRCADb \& Private & DSC=90.4\% (3D-IRCADb) & The performance drops on the proposed dataset. \\ \hline

\cite{feng2021edge} & EVS-Net & Semi-supervised segmentation of pathological liver vessels with limited labels. & Private & DSC=98.96 & Achieve a close performance of fully supervised methods with limited labeled patches \\ \hline

\cite{yang2021liver} & Inter-scale V-Net & Liver vessel segmentation based on inter-scale V-Net & 3D-IRCADb & DSC=71.6\% & The performance needs further improvement and additional investigations on other datasets are required. \\ \hline

\cite{montana2021vessel} & 2D U-Net & Vessel segmentation for automatic registration in 2D untracked LUS images & Private & DSC=64.81\% & Difficulty in segmenting rapidly varying vessel sections over time. Also, the performance needs to be improved. \\ \hline

\cite{liu2022unet} & SU-UNet & SU-UNet: A Novel Self-Updating Network for Hepatic Vessel Segmentation in CT Images & MSDC & DSC= 62.94\% & Further performance improvement and investigations on other datasets are required. \\ \hline

\cite{survarachakan2021effects} & 3D U-Net & Liver vessel segmentation by combining multiple vesselness enhancement filters. & OSLO-COMET \cite{fretland2015open} & DSC=80\% & Further investigations on other datasets are required. \\ \hline

\end{tabular}
}
\end{table}

\subsubsection{Outlook}
Many insights can be drawn from the above survey: 1) The majority of studies in the liver delineation task utilize supervised ML algorithms, especially the 2D and 3D FCN-based models as depicted in Fig.~\ref{fig:liver_map} and compared in Table~\ref{tab:liver_seg}; 2) The huge advancements in the ML field to tackle the biomedical problems. For instance, using 3D or 2.5D instead of 2D or innovating new interconnected architectures, allowing the model to understand the liver's complex structure; 3) The interconnectivity of different algorithms aiming to segment multiple tissues, showing an initiative towards creating a complete algorithm for full liver delineation; 4) The number of studies investigating tumors and vessels delineation, shown in \textcolor{black}{Fig.~\ref{fig:tumors_and_vessels_map} and summarized in Table~\ref{tab:tumor_seg} and Table~\ref{tab:vessel_seg}}, are low when compared with the ones investigating the liver's delineation, especially the vessels segmentation studies that are rare, pointing at a research area worth further investigation; 5) The severe absence of studies that tackle all liver's different tissues delineation problem. \textcolor{black}{It is worth noting that the studies that worked on solely on Private datasets have been removed from the liver segmentation comparison in Table~\ref{tab:liver_seg} and the tumors segmentation in Table~\ref{tab:tumor_seg}. Contrastingly, they were kept for the vessel segmentation comparison in Table~\ref{tab:vessel_seg} due to the scarce number of studies.}

\section{Critical Discussions}\label{sec5}
\textcolor{black}{
A wide range of DL-based algorithms have been proposed and shown state-of-the-art performance for different tasks, including automatic liver segmentation, liver tumor segmentation, and liver vessel segmentation.
The performance of DL techniques has been enhanced over time in terms of segmentation accuracy, complexity, and layers. Moreover, CNN models have been the most utilized architectures for the aforementioned segmentation tasks with 2.5D, 3D, and 4D images. Additionally, some studies have used TL, particularly fine-tuning, in which a pre-trained network (whole layers or part of them) was employed to initialize the weights. However, most DL algorithms have adopted end-to-end training without using any pre-trained models.}

\textcolor{black}{
On the other hand, it was obvious that most of the studies still have difficulties in simultaneously discriminating between tumors, inside-liver tissues, and outside-liver organs. This is in addition to the fact that extracting features that reflect the axial changes of the liver and the tumor is still challenging because of the high computational cost. This results in limited learning effects and efficiencies.
}

\begin{figure}[!ht]
\centering
\includegraphics[width=0.52\linewidth, trim={0in 0in 0in 0in}, clip]{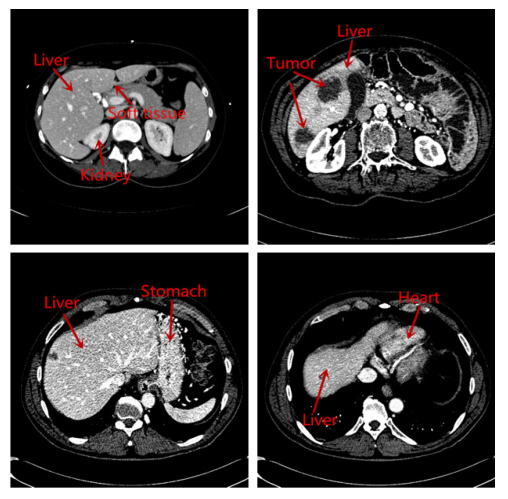}
\caption{\textcolor{black}{Some liver segmentation challenges that still impede the development of accurate semi-automatic or automatic segmentation solutions \cite{tang2020two}.}}
\label{fig:liver_tissues_challenges}
\end{figure}

\textcolor{black}{
In this context, although significant progress has been made for semi-automatic or fully automatic liver, liver tumor, and liver vessel CT image segmentation, precise and accurate segmentation is still challenging in many scenarios. For example, in the case of liver segmentation, this is due to various reasons, as visually shown in Fig.~\ref{fig:liver_tissues_challenges}. Typically, this includes: (i) the intensity similarity between the liver and its neighboring organs such as the stomach and heart); (ii) the blurring of the liver contour due to the partial volume effects; (iii) the severe pathological changes, e.g., large tumors and cirrhosis, often occur in clinical images, although its intensity is evidently not similar to that of the normal liver; and (iv) the liver shape can vary from a person to another, especially if a hepatectomy has been performed on said person.
These issues impede the segmentation of livers with complex contours and small sizes and complicate the automation of liver segmentation when diagnosing and treating CT images \cite{tang2020two}.}

\subsection{Open Challenges}
After surveying the literature and understanding the novelty of each work, we believe that the available techniques in literature, especially in the last 3 years, have advanced rapidly so that they can be soon implemented in clinical environments if that is not the case. To the best of our knowledge, we have included the key studies that are concerned with the automatic segmentation of liver, tumors, and vessels using ML algorithms. Thus, from that perspective, we see some patterns/gaps in both Fig.~\ref{fig:liver_map} and Fig.~\ref{fig:tumors_and_vessels_map} that are worth highlighting. Along our survey, we also identified gaps and challenges that need attention to be addressed.

\subsubsection{Tumors and Vessels Literature Availability}
The works utilizing ML techniques for tumors and vessel segmentation seem to be a little copious against the ones concerned with the liver. This is understandable because the advanced ML algorithms have been thoroughly investigated for the liver segmentation task first, which can then be utilized for the tumors and vessels segmentation task in a hierarchy fashion. Nonetheless, they are of scarce nature, especially the vessels segmentation task, due to the fewer available masks for hepatic vasculature needed for ML algorithms. However, with the introduction of the  MSDC-T8 dataset, the research fraternity has been boosted up to utilize ML algorithms on vessel and tumor segmentation.

\subsubsection{Classifying All Tissues}
The most prominent gap that has been identified is the absence of studies that segment all the hepatic tissues within the liver, except for \cite{Treilhard2017}. but even though the achieved results are not comparable with the state-of-the-art. One of the possible reasons for this problem might be due to the unavailability of complete masked data, including liver parenchyma, tumors, and vessels. Although the MSDC-T8 dataset provides two of the three elements required, the tumors and vessels masks, there has been another database recently introduced \cite{Tian2019} that includes the liver annotations, rendering this dataset a complete one for segmenting all the tissues within the liver.

\subsubsection{Absence of Post-operative Ground-Truth and Follow-up Datasets}
After liver resection, the surgeons/interventional radiologists are always concerned about keeping the maximum liver mass, expecting that the lost liver mass would naturally recover \cite{fausto2000liver}. However, the liver's shape is not necessarily maintained in this growth, making it irregular in the eyes of ML algorithms. When trained, they are not subjected to these unfamiliar cases, rendering these algorithms helpless in these types of situations. Thus, it is out of necessity to provide a postoperative/follow-up dataset to train the algorithms to ensure that such techniques work in the best way possible, regardless of the liver shape being delineated. Moreover, if both the pre-operative and post-operative volumes of the same patient are available, it is possible to find a mapping transformation that allows the creation of synthetic patient volumes to increase the post-operative dataset size, hence, more robust training networks.

\section{Future Directions}
\begin{figure}[!ht]
\centering
\includegraphics[width=0.75\linewidth, trim={0in 0in 0in 0in}, clip]{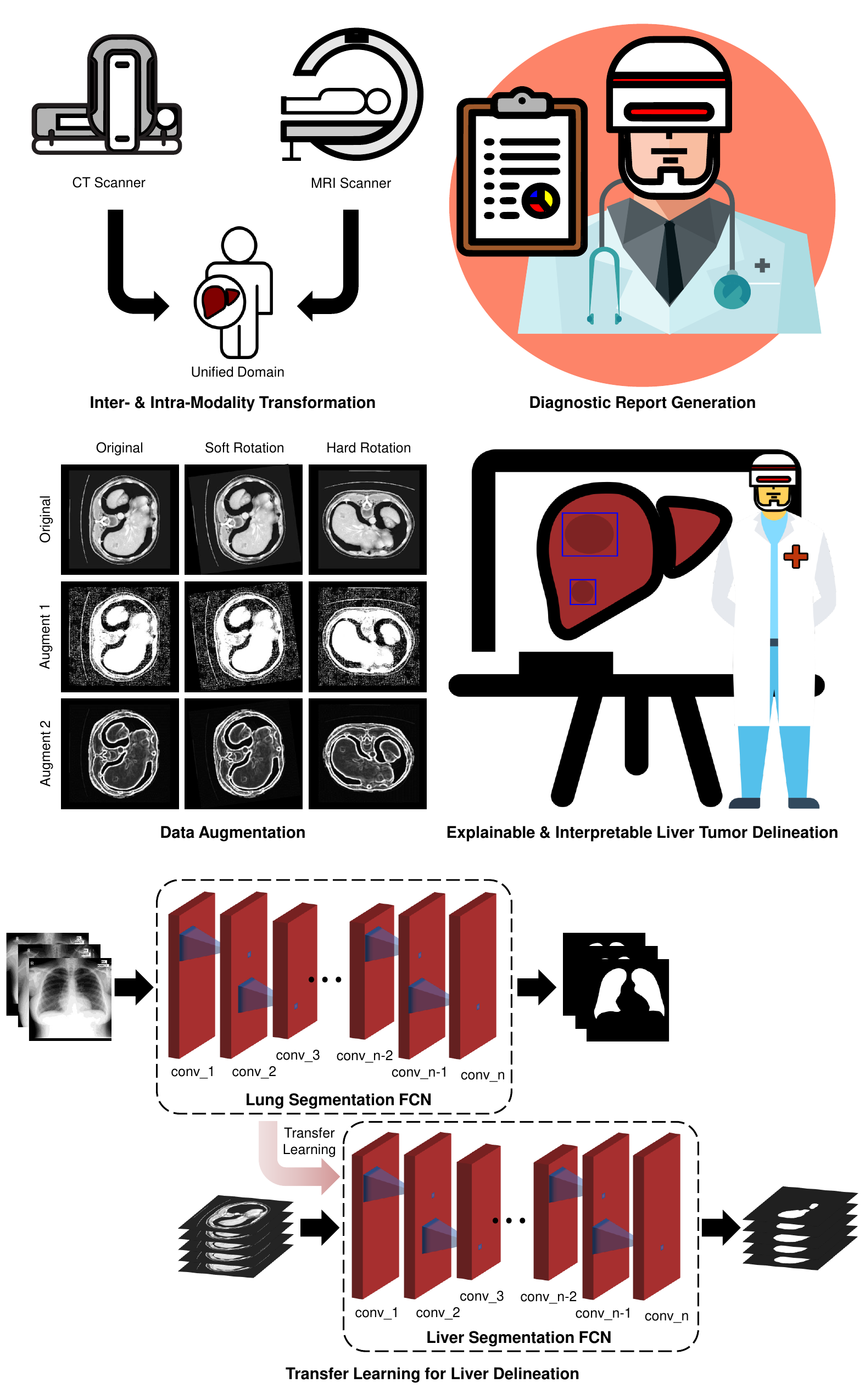}
\caption{\textcolor{black}{Reported future directions (CT scans and liver masks are taken from \cite{MSDC}, while chest X-rays and lung masks are taken from \cite{jaeger2014two}).}}
\label{future_directions}
\end{figure}

\textcolor{black}{From the surveyed literature, it is evident that a lot of aspects have been tackled; however, we compiled the below aspects that we think need more study and research, generally related to the imaging modalities, and especially related to the liver. Fig.~\ref{future_directions} highlights those aspects.}

\subsection{Inter- and Intra-Modality Transformation}
A transformation neural network should be built to transfer volumes from CT to CE-CT scans from existing datasets or create multi-phase MRI volumes from a single-phase MRI volume to allow the construction of a bigger dataset. The motivation for creating a style transfer from normal CT scans to CE-CT is that the administration of contrast agents, such as iodinated contrast medium, can increase the radiation dose that organs absorb \cite{amato2013can,sahbaee2017effect}. Similarly, a transformation from the MRI domain to the CT domain, or vice-versa, can also help in allowing MRI-based images to be fed immediately to a CT-based trained model instead of training the model on volumes from both modalities. Some works tackle the issue at hand, but further investigation can be clearly established to allow better inter- and intra-modality transformations. For instance, \cite{Jiang2018} utilized a GAN model to transfer volumes from CT to MRI, creating synthesized MRI volumes that are combined with a few real ones and training a U-Net for segmenting tumors within the lungs. Another approach that tackles the problem differently is depicted in \cite{Yang2019}. An algorithm consists of two modules where the former decomposes volumes from both MRI and CT, using VAEs and GANs, into domain-invariant content space, containing the anatomical information, and domain-specific style space, preserving the modality information. The latter module takes the common space output from the first module and inserts it into a 2D U-Net to segment the liver. 

\subsection{Data Augmentation}
Data augmentation is important in the field of deep learning for increasing the input dataset size, as the input training data is not abundant, and deep learning models are data hungry. Data augmentation can be implemented via translating the pixels, softly/harshly rotating the slices, and horizontally and/or vertically flipping the images/volumes. All these techniques are employed to increase the dataset size. Utilizing such augmentation schemes on normal images is significant as many of them can be countered in real-life scenarios. However, in the biomedical field, some of these augmentation techniques can be meaningless in the sense that flipping a slice/volume horizontally or vertically is not something that can be countered. From that point of view, a better understanding of the data augmentation techniques in the biomedical field should be attained such that the data augmentation hack is not counter-productive.

\subsection{Diagnostic Report Generation}
As ML algorithms are increasingly getting ``smarter'', it is possible to devise models that can be used in the context of diagnosis. One such piece of work \cite{Wang2020} classifies if a certain slice can be used to make a diagnostic decision. In another work, an LSTM-based language model inserts diagnostic captions with segmented tumors, which helps radiologists and surgeons in having a machine-perspective opinion of the issue at hand \cite{Tian2018}. This is important, especially in countries with enormous populations. Combining the two aforementioned works could further the technological advancements toward healthcare automation. In this way, many such smarter ways could be planned to automate the healthcare report generation and improve the healthcare system overall.

\textcolor{black}{
\subsection{Explainable and Interpretable Liver Tumor Delineation}
Explainability and interpretation are among the essential aspects when it comes to using ML and DL models in healthcare. Although ML/DL techniques have demonstrated better performance than humans in many analytical tasks, the absence of interpretations and explanations still impedes the broad utilization of these tools. Additionally, it continues to provoke critics as most ML/DL models are considered as \enquote{black-box} methods that lack interpretation \cite{amann2020explainability}. Thus, this can doubt the credibility of reached decisions and lack compelling evidence for convincing experts.
Yet, adding interpretations and explanations is not an entirely technological problem but evokes a multitude of ethical, legal, medical, and societal queries that need to be solved and explored. To that end, an increasing interest has been shown recently in developing explainable and interpretable liver tumor delineation frameworks \cite{singh2020explainable}.}

\textcolor{black}{
For instance, to improve segmentation defects in liver CT images, Mohagheghi et al. \cite{mohagheghi2022developing} develop an explainable DL boundary correction approach. The latter incorporates cascaded x-Dim models (1D and 2D), which helps in refining other models' outputs and providing robust and accurate results.
Typically, to refine incorrect segmented regions (slice-by-slice), a 2-step loop with a 1D local boundary validation scheme is firstly implemented before applying a 2D image patch segmentation approach. 
In \cite{turco2022interpretable}, an interpretable ML approach that characterizes focal liver lesions by contrast-enhanced ultrasound (CEUS) is proposed. It helps differentiate between benign and malignant focal liver lesions (FLLs) on CEUS.
Typically, it relies on defining the ROIs, extracting spatiotemporal features, filtering for dimensionality reduction using PCA, and applying different classifiers including, k-nearest neighbor (kNN), random forest (RF), support vector machine (SVM), logistic regression (LR), and soft voting classifier (sVC).
Moving forward, an explainable liver tumor delineation in surgical specimens is proposed in \cite{zhang2021explainable}. Specifically, hyperspectral imaging and a multi-task U-Net model are deployed to achieve an overall sensitivity of 94.48\%, outperforming SVM by a large margin.
}

\textcolor{black}{
\subsection{Transfer Learning (TL) for Liver Delineation}
It has been demonstrated in many studies, such as \cite{prasad2021modifying,conze2021abdominal,liu2022free}, that most DL models can not segment accurately liver tissues if (i) small training datasets are considered; and/or (ii) there is a discrepancy or data distribution inconsistency between training and test data \cite{zoetmulder2022domain}.
To that end, TL has recently received increasing attention due to its ability to (i) provide high-quality decision support; (ii) require less training data compared to conventional DL algorithms; and (iii) reduce the domain shift between target domain data (test) and source domain data (training).
Typically, these properties came from the fact that TL models are already trained on large generic datasets. Thus, only a task-specific dataset is required to customize these models. Consequently, the need to train TL models from scratch is eliminated. 
TL is based either on fine-tuning pre-trained models or DA. The former is performed by reusing previously fully trained networks with a specific dataset for a particular purpose \cite{karimi2021transfer,nowak2021detection}. By contrast, DA is accomplished to tackle domain shifts often occur when the medical images are recorded by different equipment or in different environments \cite{hong2022source}. 
DA is mainly based on different approaches: (i) divergence-based DA; (ii) adversarial-based DA using GANs \cite{hong2022unsupervised}; and (iii) reconstruction-based DA using GANs or stacked autoencoders (SAEs) or GANs \cite{yao2022novel}.
}

\textcolor{black}{
On the other hand, labeling liver/tumor segmentation images in the target domain is a crucial challenge, which necessitates the intervention of experienced radiologists \cite{yao2022novel}. Fortunately, unsupervised DA techniques help transfer knowledge across domains without the need for annotated data in the target domain. Hence, they can be beneficial in alleviating the data labeling process, which is costly and time-consuming. 
For instance, an unsupervised DA scheme cross-modality liver segmentation using self-learning and joint adversarial learning is proposed in \cite{yao2022novel}. Typically, a post-situ identification strategy along with a shape-entropy-aware and joint semantic-aware adversarial learning are introduced to implicitly align the distributions of task-related characteristics derived from the target and source domains. Moving on, Hong et al. \cite{hong2022source} introduce cross-modality abdominal multi-organ segmentation using source-free unsupervised DA. Similarly, a DA-based liver tumor segmentation scheme using adversarial learning in multi-phase CT images is presented in \cite{jain2022unsupervised}.
}

\section{Conclusions}\label{sec6}
We have created a survey covering the major studies available between 2014 and 2022 employing automatic ML algorithms on the liver, tumors, and/or vessel segmentation. \textcolor{black}{Three tables are created to summarize and compare the different studies that delved into the challenges of the liver, tumors, and vessel segmentations.} The following conclusions are drawn: 1) We summarize the existence of full liver, tumor, and vessels manual labeled datasets, highlighting that the MSDC-T8 is the most complete one; 2) For the liver segmentation task, the biggest cluster of ML algorithms fall under the 2D FCN umbrella; 3) Similarly, for the tumors segmentation task, 2D FCNs constituted the majority of ML algorithms used; 4) The common algorithms shared between the liver and tumors segmentation tasks were also highlighted; 5) Even though it is of high importance, few works have addressed the vessels segmentation task using ML techniques; 6) We identified numerous challenges and future directions for the researchers to address and improve, where the most prominent one is the absence of studies that classify all the associated tissues within the liver.

In the end, we believe that this survey is a concise and comprehensive reference that could assist the research fraternity in the field of automatic ML-based liver tissues segmentation techniques. We also believe that an in-depth analysis could be further investigated, focusing on the selection of a particular ML algorithm for a specific task.

\section*{Acknowledgment}
This publication was made possible by an Award [GSRA6-2-0521-19034] from Qatar National Research Fund (a member of Qatar Foundation). The contents herein are solely the responsibility of the authors.

\bibliography{references}

\end{document}